%% file: Probabilistic_Occupancy_Grid_for_Radio_Based_SLAM.tex
\documentclass[journal]{IEEEtran}
\usepackage[table]{xcolor}

\newcommand{\exportFigures}{true} 
\input{./definitions.tex}
\newcommand{\tikzfolder}{./compiledPlots/}
\input{pgfplot}

\IEEEoverridecommandlockouts

\begin{document}
	\allowdisplaybreaks
	\frenchspacing
	\title{\huge Probabilistic Occupancy Grid for Radio-Based SLAM}
	\author{
		\IEEEauthorblockN{
			Xuhong~Li$^\ast$, 
			Erik~Leitinger$^\dagger$,
			Fredrik Tufvesson$^\ast$,
			and Florian~Meyer$^\ddagger$}\\
		\vspace*{1mm}
		\IEEEauthorblockA{$^\ast$Department of Electrical and Information Technology, Lund University, Sweden. \\ 
			$^\dagger$Institute of Comm. Networks and Satellite Comms., Graz University of Technology, Austria \\ 
			$^\ddagger$Department of Electrical and Computer Engineering, University of California San Diego, USA\\ 
			{\small Email: \{xuhong.li, fredrik.tufvesson\}@eit.lth.se, erik.leitinger@tugraz.at, flmeyer@ucsd.edu} 
			\thanks{This work was supported by the Knut and Alice Wallenberg Foundation under WASP Scholarship Program (KAW 2023.0509), by the Strategic Research Area Excellence Center at Linköping–Lund in Information Technology (ELLIIT), by NextG2Com funded by the VINNOVA program for Advanced Digitalisation with grant number 2023-00541, by the Austrian Research Promotion Agency (FFG) under the PRISM project (620753), by the European Union’s Horizon Europe research and innovation program under Grant 101192113, and by the National Science Foundation (NSF) under CAREER Award No.~2146261.}
		}
		\vspace*{-7mm}
	}

	\maketitle
	\renewcommand{\baselinestretch}{0.98}\small\normalsize
	\renewcommand{\baselinestretch}{1} 
	
	\begin{abstract}
		\subfile{./InputFiles/abstract}
	\end{abstract}
	
	\IEEEpeerreviewmaketitle
	\acresetall
	
	\vspace{-2mm}
	\section{Introduction}
	\subfile{./InputFiles/Introduction}
	
	\vspace{-1mm}
	\section{Geometrical Relations}
	\subfile{./InputFiles/GeometricalRelations}

	\vspace{-1mm}
	\section{System Model Problem Formulation }
	\subfile{./InputFiles/SystemModel}

	\section{Results}
	\subfile{./InputFiles/Results}

	\vspace{-3mm}
	\section{Conclusion}
	\subfile{./InputFiles/Conclusion}

	
	\renewcommand{\baselinestretch}{0.95}\small\normalsize
	\bibliographystyle{IEEEtran}
	\balance
	\bibliography{IEEEabrv,./references}

\end{document}

%% file: definitions.tex
\usepackage{tikz}
\usetikzlibrary{backgrounds}
\tikzstyle{load}   = [ultra thick,-latex]
\tikzstyle{stress} = [-latex]
\tikzstyle{dim}    = [latex-latex]
\tikzstyle{axis}   = [-latex,black!55]

\definecolor{green(pigment)}{rgb}{0.0, 0.65, 0.31}
\definecolor{frenchblue}{rgb}{0.0, 0.45, 0.73} 
\definecolor{mediumcandyapplered}{rgb}{0.89, 0.02, 0.17}
\definecolor{darkGreyGreen598081}{rgb}{0.349, 0.502, 0.506}
\definecolor{darkGreyGreen2b6a99}{rgb}{0.349, 0.502, 0.506}
\definecolor{darkBlue1b7c3d}{rgb}{0.14, 0.35, 0.51}

\definecolor{redb11927}{rgb}{0.734, 0.104, 0.162}
\definecolor{blue4695be}{rgb}{0.171, 0.364, 0.465}

\usepackage{tkz-euclide}
\usetikzlibrary{arrows.meta}
\usepackage{setspace}
\usepackage{bm}
\usepackage{mathtools}
\usepackage{graphicx}
\usepackage{relsize}
\usepackage{cite}
\usepackage{amsmath}
\usepackage{color}
\usepackage{psfrag}
\usepackage{amssymb}
\usepackage{dblfloatfix}
\usepackage[font=small]{caption}
\usepackage{tabularx}
\usepackage{makecell}
\usepackage{lipsum}
\usepackage{float}
\usepackage{subfiles}
\usepackage{epstopdf}
\usepackage{verbatim}
\usepackage{amsmath}
\usepackage{ifthen}
\usepackage{pgfplots}
\usepackage{pgfplotstable}
\usepackage{tikzscale}
\usepackage{acronym}
\usepackage{amsfonts}
\usepackage{tikz-3dplot}
\usepackage{cuted}
\usepackage{subfig}
\usepackage{balance}

\usetikzlibrary{intersections}   
\usepgfplotslibrary{fillbetween}

\usetikzlibrary{spy,backgrounds}
\usetikzlibrary{fit}
\usetikzlibrary{calc}

\usetikzlibrary{3d}
\usetikzlibrary{shapes}
\usetikzlibrary {patterns,patterns.meta}
\usepackage{pgffor}
\usetikzlibrary{shapes.geometric}
\usetikzlibrary{angles,quotes}

\renewcommand{\baselinestretch}{1} 

\colorlet{veccol}{green!50!black}
\colorlet{projcol}{blue!70!black}
\colorlet{myblue}{blue!80!black}
\colorlet{myred}{red!90!black}
\colorlet{mydarkblue}{blue!50!black}
\tikzset{>=latex} 
\tikzstyle{proj}=[projcol!80,line width=0.08] 
\tikzstyle{area}=[draw=veccol,fill=veccol!80,fill opacity=0.6]
\tikzstyle{vector}=[-stealth,myblue,thick,line cap=round]
\tikzstyle{unit vector}=[->,veccol,thick,line cap=round]
\tikzstyle{dark unit vector}=[unit vector,veccol!70!black]
\usetikzlibrary{angles,quotes} 

\definecolor{RDlightgreen}{RGB}{141 192 69}
\definecolor{RDgreen}{rgb}{0.3647, 0.4275, 0.2667}
\definecolor{RDdarkgreen}{rgb}{0.2196, 0.2196, 0.2196}
\definecolor{RDmaroon}{rgb}{.522,.22,.353} %

\definecolor{IEEEblue}{RGB}{0 98 155}
\definecolor{IEEElightblue}{RGB}{0 181 226}
\definecolor{IEEEturquoise}{RGB}{0 156 166}
\definecolor{IEEEred}{RGB}{186 12 47}
\definecolor{IEEEgreen}{RGB}{0 132 61}
\definecolor{IEEElightgreen}{RGB}{120 190 32}
\definecolor{IEEEorange}{RGB}{225 163 0}
\definecolor{IEEEyellow}{RGB}{255 209 0}
\definecolor{IEEEviolett}{RGB}{152 29 151}
\definecolor{IEEEdarkmaroon}{RGB}{134 31 65}
\definecolor{IEEEdarkorange}{RGB}{232 119 34}

\def\sfv{{sfv}}

\def\ip{{IP}}

\newcommand{\ist}{\hspace*{.3mm}}
\newcommand{\rmv}{\hspace*{-.3mm}}
\newcommand{\iist}{\hspace*{1mm}}
\newcommand{\rrmv}{\hspace*{-1mm}}
\newcommand{\nn}{\nonumber}

\DeclareMathAlphabet{\mathsfbr}{OT1}{cmss}{m}{n}
\SetMathAlphabet{\mathsfbr}{bold}{OT1}{cmss}{bx}{n}
\DeclareRobustCommand{\msf}[1]{%
	\ifcat\noexpand#1\relax\msfgreek{#1}\else\mathsfbr{#1}\fi
}

\makeatletter
\newcommand{\msfgreek}[1]{\csname s\expandafter\@gobble\string#1\endcsname}
\makeatother

\DeclareRobustCommand{\mcal}[1]{%
	\ifcat\noexpand#1\relax\mathnormal{#1}\else\cal{#1}\fi
}
\DeclareRobustCommand{\BM}[1]{%
	\ifcat\noexpand#1\relax\bm{\boldUppercaseItalicGreek{#1}}\else\bm{#1}\fi
}
\makeatletter
\newcommand{\boldUppercaseItalicGreek}[1]{\csname var\expandafter\@gobble\string#1\endcsname}
\makeatother

\newcommand{\rv}[1]{\msf{#1}}
\newcommand{\RV}[1]{\bm{\msf{#1}}}

\newcommand{\V}[1]{\bm{#1}}

\let\geq\geqslant

\DeclareMathVersion{timesmath}
\SetSymbolFont{letters}{timesmath}{OML}{ntxmi}{m}{it}
\DeclareMathVersion{timesmathbold}
\SetSymbolFont{letters}{timesmathbold}{OML}{ntxmi}{b}{it}

\makeatletter
\newif\ifAC@uppercase@first%
\def\Aclp#1{\AC@uppercase@firsttrue\aclp{#1}\AC@uppercase@firstfalse}%
\def\AC@aclp#1{%
	\ifcsname fn@#1@PL\endcsname%
	\ifAC@uppercase@first%
	\expandafter\expandafter\expandafter\MakeUppercase\csname fn@#1@PL\endcsname%
	\else%
	\csname fn@#1@PL\endcsname%
	\fi%
	\else%
	\AC@acl{#1}s%
	\fi%
}%
\def\Acp#1{\AC@uppercase@firsttrue\acp{#1}\AC@uppercase@firstfalse}%
\def\AC@acp#1{%
	\ifcsname fn@#1@PL\endcsname%
	\ifAC@uppercase@first%
	\expandafter\expandafter\expandafter\MakeUppercase\csname fn@#1@PL\endcsname%
	\else%
	\csname fn@#1@PL\endcsname%
	\fi%
	\else%
	\AC@ac{#1}s%
	\fi%
}%
\def\Acfp#1{\AC@uppercase@firsttrue\acfp{#1}\AC@uppercase@firstfalse}%
\def\AC@acfp#1{%
	\ifcsname fn@#1@PL\endcsname%
	\ifAC@uppercase@first%
	\expandafter\expandafter\expandafter\MakeUppercase\csname fn@#1@PL\endcsname%
	\else%
	\csname fn@#1@PL\endcsname%
	\fi%
	\else%
	\AC@acf{#1}s%
	\fi%
}%
\def\Acsp#1{\AC@uppercase@firsttrue\acsp{#1}\AC@uppercase@firstfalse}%
\def\AC@acsp#1{%
	\ifcsname fn@#1@PL\endcsname%
	\ifAC@uppercase@first%
	\expandafter\expandafter\expandafter\MakeUppercase\csname fn@#1@PL\endcsname%
	\else%
	\csname fn@#1@PL\endcsname%
	\fi%
	\else%
	\AC@acs{#1}s%
	\fi%
}%
\edef\AC@uppercase@write{\string\ifAC@uppercase@first\string\expandafter\string\MakeUppercase\string\fi\space}%
\def\AC@acrodef#1[#2]#3{%
	\@bsphack%
	\protected@write\@auxout{}{%
		\string\newacro{#1}[#2]{\AC@uppercase@write #3}%
	}\@esphack%
}%
\def\Acl#1{\AC@uppercase@firsttrue\acl{#1}\AC@uppercase@firstfalse}
\def\Acf#1{\AC@uppercase@firsttrue\acf{#1}\AC@uppercase@firstfalse}
\def\Ac#1{\AC@uppercase@firsttrue\ac{#1}\AC@uppercase@firstfalse}
\def\Acs#1{\AC@uppercase@firsttrue\acs{#1}\AC@uppercase@firstfalse}
\robustify\Aclp
\robustify\Acfp
\robustify\Acp
\robustify\Acsp
\robustify\Acl
\robustify\Acf
\robustify\Ac
\robustify\Acs
\makeatother

\makeatletter
\def\underbracex#1#2{\mathop{\vtop{\m@th\ialign{##\crcr
				$\hfil\displaystyle{#2}\hfil$\crcr
				\noalign{\kern3\p@\nointerlineskip}%
				#1\crcr\noalign{\kern3\p@}}}}\limits}

\def\underbracea{\underbracex\upbracefilla}

\def\upbracefilla{$\m@th \setbox\z@\hbox{$\braceld$}%
	\bracelu\leaders\vrule \@height\ht\z@ \@depth\z@\hfill 
	\leaders\vrule \@height\ht\z@ \@depth\z@\hfill\bracerd
	\braceld\leaders\vrule \@height\ht\z@ \@depth\z@\hfill
	\kern\p@\vrule \@width\p@\kern\p@\vrule \@width\p@\kern\p@\vrule \@width\p@ 
	$}

\def\underbraceaa{\underbracex\upbracefillaa}

\def\upbracefillaa{$\m@th \setbox\z@\hbox{$\braceld$}%
	\bracelu\leaders\vrule \@height\ht\z@ \@depth\z@\hfill 
	\kern\p@\vrule \@width\p@\kern\p@\vrule \@width\p@\kern\p@\vrule \@width\p@
	$}

\def\underbraceb{\underbracex\upbracefillb}

\def\upbracefillb{$\m@th \setbox\z@\hbox{$\braceld$}%
	\vrule \@width\p@\kern\p@\vrule \@width\p@\kern\p@\vrule \@width\p@\kern\p@
	\leaders\vrule \@height\ht\z@ \@depth\z@\hfill\bracerd
	\braceld\leaders\vrule \@height\ht\z@ \@depth\z@\hfill
	\kern\p@\vrule \@width\p@\kern\p@\vrule \@width\p@\kern\p@\vrule \@width\p@
	$}

\def\upbracefillc{$\m@th \setbox\z@\hbox{$\braceld$}%
	\vrule \@width\p@\kern\p@\vrule \@width\p@\kern\p@\vrule \@width\p@\kern\p@
	\leaders\vrule \@height\ht\z@ \@depth\z@\hfill
	\kern\p@\vrule \@width\p@\kern\p@\vrule \@width\p@\kern\p@\vrule \@width\p@
	$}

\def\underbraced{\underbracex\upbracefilld}
\def\upbracefilld{$\m@th \setbox\z@\hbox{$\braceld$}%
	\vrule \@width\p@\kern\p@\vrule \@width\p@\kern\p@\vrule \@width\p@\kern\p@
	\leaders\vrule \@height\ht\z@ \@depth\z@\hfill\braceru$}

\makeatother

\acrodef{2d}[2D]{two-dimensional}
\acrodef{3d}[3D]{three-dimensional}
\acrodef{5g}[5G]{5th-generation}
\acrodef{pnt}[PNT]{positioning, navigation and timing}
\acrodef{pa}[PA]{physical anchor}
\acrodef{bs}[BS]{base station}
\acrodef{awgn}[AWGN]{additive white Gaussian noise}
\acrodef{va}[VA]{virtual anchor}
\acrodef{mpc}[MPC]{multipath component}
\acrodef{fa}[FA]{false alarm}
\acrodef{far}[FAR]{false alarm rate}
\acrodef{pva}[PR]{potential ray}
\acrodef{mf}[MF]{map feature}
\acrodef{pmf}[PMF]{potential \ac{mf}}
\acrodef{smc}[SMC]{specular multipath component}
\acrodef{psmc}[PSMC]{potential \ac{smc}}
\acrodef{sfv}[SFV]{surface feature vector}
\acrodef{ip}[IP]{interaction point}
\acrodef{psfv}[PSFV]{potential \ac{sfv}}
\acrodef{slam}[SLAM]{simultaneous localization and mapping}
\acrodef{Mp}[MP]{Multipath-based}
\acrodef{mpslam}[MP-SLAM]{multipath-based \ac{slam}}
\acrodef{pmf}[PMF]{probability mass function}
\acrodef{pdf}[PDF]{probability density function}
\acrodef{cdf}[CDF]{cummulative distribution function}
\acrodef{bp}[BP]{belief propagation}
\acrodef{spa}[SPA]{sum-product algorithm}
\acrodef{fg}[FG]{factor graph}
\acrodef{mmse}[MMSE]{minimum mean-square error}
\acrodef{simo}[SIMO]{single-input-multiple-output}
\acrodef{miso}[MISO]{multiple-input-single-output}
\acrodef{mimo}[MIMO]{multiple-input-multiple-output}
\acrodef{dmimo}[D-MIMO]{distributed multiple-input-multiple-output}
\acrodef{mmwave}[mmWave]{millimeter-wave}
\acrodef{ue}[UE]{mobile user}
\acrodef{los}[LoS]{line-of-sight}
\acrodef{nlos}[NLoS]{non \ac{los}}
\acrodef{olos}[OLoS]{obstructed \ac{los}}
\acrodef{aoa}[AoA]{angle-of-arrival}
\acrodef{aod}[AoD]{angle-of-departure}
\acrodef{roi}[RoI]{region-of-interest}
\acrodef{snr}[SNR]{signal-to-noise ratio}
\acrodef{ospa}[OSPA]{optimal subpattern assignment}
\acrodef{mospa}[MOSPA]{mean \ac{ospa}}
\acrodef{dmc}[DMC]{dense multipath component}
\acrodef{sr}[SR]{super-resolution}
\acrodef{sbl}[SBL]{sparse Bayesian learning}
\acrodef{sde}[SDE]{sequential detection and estimation}
\acrodef{kest}[KEST]{Kalman enhanced super resolution tracking}
\acrodef{std}[STD]{standard deviation}
\acrodef{da}[DA]{data association}
\acrodef{fim}[FIM]{Fisher information matrix}
\acrodef{crlb}[CRLB]{Cram\'{e}r–-Rao lower bound}
\acrodef{pcrlb}[PCRLB]{posterior Cram\'{e}r–-Rao lower bound}
\acrodef{meb}[MEB]{mapping error bound}
\acrodef{peb}[PEB]{position error bound}
\acrodef{oeb}[OEB]{orientation error bound}
\acrodef{va}[VA]{virtual anchor}
\acrodef{eadf}[EADF]{effective aperture distribution function}
\acrodef{lhf}[LHF]{likelihood function}
\acrodef{rmse}[RMSE]{root mean square error} 
\acrodef{ps}[PS]{point scatterer}
\acrodef{imm}[IMM]{interacting multiple model}
\acrodef{rf}[RF]{radio frequency}
\acrodef{iid}[iid]{independent and identically distributed}
\acrodef{prop}[PROP]{proposed algorithm}
\acrodef{refa}[REFA]{reference algorithm}
\acrodef{rt}[RT]{ray tracing}
\acrodef{ceda}[CEDA]{snapshot-based parametric channel estimation and detection algorithm}

%% file: pgfplot.tex
\usetikzlibrary{backgrounds}
\usetikzlibrary{calc,arrows.meta}
\ifthenelse{\equal{\exportFigures}{true}}
{
  \usepgfplotslibrary{external}\tikzexternalize[prefix=\tikzfolder]
  \tikzset{external/system call={pdflatex \tikzexternalcheckshellescape -halt-on-error -interaction=batchmode -jobname "\image" "\texsource"}}

}
{}

\def\addlegendimage{\csname pgfplots@addlegendimage\endcsname}
\def\addlegendentry{\csname pgfplots@addlegendentry\endcsname}

\newcommand{\pgfref}[1]
{\ifthenelse{\equal{\exportFigures}{true}}
{\tikzexternaldisable\ref{#1}\tikzexternalenable}
{\ref{#1}}}

\pgfplotsset{compat=newest} 

\newlength\figureheight 
\newlength\figurewidth 
\usetikzlibrary{patterns,arrows}
\usepgfplotslibrary{colormaps}
\usetikzlibrary{plotmarks}
\pgfplotsset{every axis/.append style={
  label style={font=\normalsize},
  tick label style={font=\scriptsize},
  xticklabel={
   \ifdim \tick pt < 0pt
    \pgfmathparse{abs(\tick)}%
    \llap{$-{}$}\pgfmathprintnumber{\pgfmathresult}
   \else
    \pgfmathprintnumber{\tick}
   \fi}
   }}
\usetikzlibrary{decorations.markings}
\makeatletter
\tikzset{
  nomorepostactions/.code={\let\tikz@postactions=\pgfutil@empty},
  mymarkfixednumber/.style n args={3}{decoration={markings,
    mark= between positions 0.1 and 1 step (1/#3)*\pgfdecoratedpathlength with{%
        \tikzset{#2,every mark}\tikz@options
        \pgfuseplotmark{#1}%
      },  
    },
    postaction={decorate},
    /pgfplots/legend image post style={
        mark=#1,mark options={#2},every path/.append style={nomorepostactions}
    },
  },
}
\tikzset{
  nomorepostactions/.code={\let\tikz@postactions=\pgfutil@empty},
  mymarkfixeddistance/.style n args={3}{decoration={markings,
    mark= between positions 0 and 1 step #3cm with{%
        \tikzset{#2,every mark}\tikz@options
        \pgftransformresetnontranslations%
        \pgfuseplotmark{#1}%
      },  
    },
    postaction={decorate},
    /pgfplots/legend image post style={
        mark=#1,mark options={#2},every path/.append style={nomorepostactions}
    },
  },
}
\tikzset{
  nomorepostactions/.code={\let\tikz@postactions=\pgfutil@empty},
  mymark/.style n args={3}{decoration={markings,
    mark= between positions 0 and 1 step 0.75cm with{%
        \tikzset{#2,every mark}\tikz@options
        \pgftransformresetnontranslations%
        \pgfuseplotmark{#1}%
      },  
    },
    postaction={decorate},
    /pgfplots/legend image post style={
        mark=#1,mark options={#2},every path/.append style={nomorepostactions}
    },
  },
}
\makeatother

\definecolor{colA}{rgb}{0,0,1}
\definecolor{colB}{rgb}{1.00000,0.0,0.00000}
\definecolor{colC}{rgb}{0.1333,0.5451,0.1333}
\definecolor{colD}{rgb}{0.9,0.0,0.9}
\definecolor{colE}{rgb}{0,1,1}
\definecolor{colF}{rgb}{1.00000,0.55,0.00000}

\def\mymarksize{1.5}
\def\mymarksize2{2.5}

\pgfdeclareplotmark{markarrow}
{%
  \pgfpathmoveto{\pgfqpoint{\pgfplotmarksize}{0pt}}
  \pgfpathlineto{\pgfqpointpolar{130}{1.5\pgfplotmarksize}}
  \pgfpathmoveto{\pgfqpoint{\pgfplotmarksize}{0pt}}
  \pgfpathlineto{\pgfqpointpolar{-130}{1.5\pgfplotmarksize}}
  \pgfusepathqstroke
}

%% file: InputFiles/abstract.tex
Sensing is an integral part of 6G and beyond systems, providing exceptional environmental perception along with communication. \Ac{rf}--based sensing often relies on simplified geometric assumptions (e.g., point scatterers or planar surfaces) to model specular multipath and keep inference tractable. However, such representations are limited in their ability to capture extended objects with complex geometries and properties. This paper presents a probabilistic occupancy grid framework for radio-based \ac{slam}, jointly reconstructing geometric structures and their RF-related properties. The proposed occupancy grid map representation is integrated into a multipath-based \ac{slam} formulation to enable simultaneous mobile-agent localization and environment mapping using multipath measurements.

To connect \ac{rf} measurements with the grid map, a surface model is employed to describe candidate reflection paths, while occupancy grid cell states capture measurement uncertainties and fine--grained geometric details. \ac{rf}--related object properties are represented through reflection coefficients. The proposed framework offers a principled, proof-of-concept approach to physically interpretable radio-based mapping, and simulation results demonstrate accurate reconstruction of geometry and material properties, as well as high-accuracy localization. In addition, the results highlight the potential to use prior occupancy maps obtained from other radio devices or complementary sensors for subsequent map extension and refinement\vspace{-2mm}.

%% file: InputFiles/Introduction.tex
\begin{figure}[!t]
	\centering
	\vspace*{-3mm}
	\hspace*{1.5mm}\subfloat[\label{subfig:MIMOenv}]
	{\hspace*{-3.5mm}\includegraphics[scale=1]{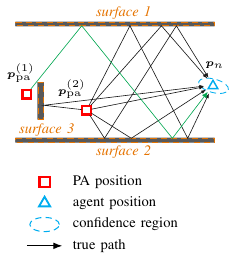}}
	\hspace*{0mm}\subfloat[\label{subfig:RayDemoSinglePath}]
	{\hspace*{-6mm}\includegraphics[scale=1]{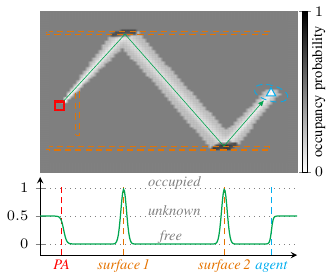}}
	\caption{(a) Geometrical depiction of a D-MIMO radio propagation environment consisting of three reflective surfaces, two PAs and a mobile agent. (b) Illustration of the interaction between grid cells and the RF double-bounce path (green) from PA\,$ 1 $ at position $ \V{p}_{\mathrm{pa}}^{(1)} $ to the mobile agent at $ \V{p}_{n} $, accounting for measurement and agent position uncertainties. The resulting occupancy probability profile is also shown, where traversed cells converge toward \textit{free} state, surface-hit cells toward \textit{occupied} state, and all other cells remain in an \textit{unknown} (unexplored) state.}
	\label{fig:FactorGraphSum}
\end{figure}

Location and environment awareness are integrated capabilities of future 6G communication networks, supporting applications such as location-aware security, autonomous systems, and digital twins in diverse scenarios, notably indoor and urban environments. In such environments, dense available infrastructure facilitates radio-based sensing \cite{GonFurKalValDarSheSheBayWymProcIEEE2024}. However, the impact of complex geometries with heterogeneous electromagnetic properties on \ac{rf} signal propagation cannot be sufficiently captured with simple parametric geometric models (e.g., \acp{va}, scatterers or smooth surfaces) commonly used in existing mapping methods \cite{Erik_SLAM_TWC2019,KimGranSveKimWym:TVT2022, LiLeiCaiTuf:ICC2024, GeWymSev:TSP2025,LiangTSP2025}, motivating the need for more realistic and physically informed environment mapping methods.

For perception of complex environments, a well-established approach is occupancy grid mapping, which discretizes the environment into fine spatial cells, each associated with a random variable encoding the probability of occupancy. This representation also facilitates multi-sensor fusion by providing a shared spatial frame that supports consistent map updates over time. Grid mapping is widely used in robotics for applications such as navigation and \ac{slam} \cite{Thrun_OGM2003, Dhiman_ICRA2014, Dominik_RobJour2018, Kwon_TRob2019, shankar_mrfmap2020, Robbiano_TITS2022}, where direct or single-bounce rays and independence between grid cell states are commonly assumed to maintain inference tractability. However, its application to radio-based mapping \cite{IrishICRA2014} remains limited given substantial computational and modeling challenges. Geometrically, it is difficult to identify how \ac{rf} signals interact with grid cells, especially in strong multipath propagation scenarios where higher-order bounce paths can contribute substantially to the received \ac{rf} signals and convey rich structural information \cite{SteGanMeiLeiWitZemPed:TAP2016,LiJiaWymWen:WCL2023}. It is also necessary to consider the statistical dependence between occupancy states of grid cells, when they are traversed or hit by the same or jointly influenced by multiple paths from one or more devices. In addition, jointly inferring grid occupancy states and mobile agent state within a unified Bayesian framework that encodes time-varying radio channel information and uncertainties is challenging.

Feature-based \ac{mpslam} methods \cite{Erik_SLAM_TWC2019,KimGranSveKimWym:TVT2022, LiLeiCaiTuf:ICC2024, GeWymSev:TSP2025,LiangTSP2025} provide the principled foundation for the integration of the occupancy grid. \ac{mpslam} exploiting \ac{va} or surface models has proven reliable in modeling and estimating dominant \ac{rf} propagation paths such as \ac{los} paths and specular reflections. In particular, \cite{LeiVenTeaMey:TSP2023, MVASLAM_TWC2025Arxiv} have demonstrated reliable ray tracing with up to double-bounce paths using adaptive data fusion. These feature models can be integrated into a hierarchical geometric representation within \ac{mpslam} to explain possible \ac{rf} signal--environment interaction processes, while occupancy-grid cell states capture fine map details. In addition, the \ac{mpslam} problem can be efficiently formulated as a factorized posterior on a factor graph representation \cite{Erik_SLAM_TWC2019,LeiVenTeaMey:TSP2023}, and approximate inference is performed applying the message passing rules of the \ac{spa} \cite{FG_SPA_TIT2001,Loeliger2004SPM}. This structure yields highly scalable algorithms and naturally accommodates extensions like occupancy grid mapping and supports adaptive data fusion.

\textit{Contribution:} In this work, we introduce a Bayesian \ac{mpslam} algorithm that jointly localizes the mobile agent and estimates the occupancy states of grid cells, allowing the mapping of complex environmental geometries using \ac{dmimo} systems \cite{SavicVTC2015, FacDeuKesWilColWitLeiSecWym:ICC2023,FasDeuKesWilColWitLeiSecWym:STSP2025,TenWymKesDeySveJCS2026}. Building on our previous work of \ac{mpslam} \cite{LeiVenTeaMey:TSP2023, MVASLAM_TWC2025Arxiv}, we establish a hierarchical geometric model that involves (i) \acp{sfv} modeling reflective surfaces and associated signal attenuation factors and (ii) occupancy grid cells. With \acp{sfv}, we resolve candidate reflected paths up to double bounces. The Bayesian estimation of the agent state, \ac{sfv} states, and cell occupancy states is established by applying the message passing algorithm to the factor graph representation of the proposed statistical model. We exploit a forward sensor model that considers the statistical dependence between grid cells and jointly updates their occupancy states after fusing information across detected propagation paths and \acp{bs}. Measurement variances that capture uncertainties in range, angle, amplitude, and related parameters are characterized through amplitude statistics using Fisher information, thereby refining the measurement likelihoods and enabling rich channel information to be encoded into grid states. Using synthetic data, this proof-of-concept study demonstrates (i) accurate localization and mapping performance and (ii) the ability to continuously expand and update an established map.

%% file: InputFiles/GeometricalRelations.tex
We consider a \ac{dmimo} system operating in a \ac{2d} dynamic scenario with horizontal-only signal propagation. At each discrete time $ n $, $ J $ distributed static \acp{pa} (e.g., \acp{bs}) at known positions $\V{p}_{\mathrm{pa}}^{(j)} =[p_{\mathrm{pa},\mathrm{x}}^{(j)} \iist p_{\mathrm{pa},\mathrm{y}}^{(j)}]^{\mathrm{T}} \rmv\in\rmv \mathbb{R}^{2\rmv\times\rmv1} $ and $j \in \{1,\dots,J\}$ transmit \ac{rf} signals, which are received by the mobile agent at unknown and time-varying position $\V{p}_{n} =[p_{n,\mathrm{x}} \iist p_{n,\mathrm{y}}]^{\mathrm{T}} \rmv\in\rmv \mathbb{R}^{2\rmv\times\rmv1} $. Each \ac{pa} is equipped with a $ H_{\mathrm{tx}} $-element antenna array with known orientation $\Delta\phi^{(j)}$, and the mobile agent is equipped with a $ H_{\mathrm{rx}} $-element antenna array with unknown orientation $\Delta\varphi_{n}$.

In the \ac{rf} propagation environment (i.e., \ac{roi}), there exist $ S $ static reflective surfaces, which are indexed by $s \in \mathcal{S} \triangleq \{1,\dots,S\} $ and modeled by \acp{sfv} at positions $ \V{p}_{s,\mathrm{\sfv}} \rmv\in\rmv \mathbb{R}^{2\rmv\times\rmv1}$ denoting the mirror images of the origin $[0,0]^\mathrm{T} $ on the surfaces. We consider specular multipath components with up to two reflections. Each of the reflected paths can be modeled by a \ac{va} at position $ \V{p}_{ss',\mathrm{va}}^{(j)} \rmv\in\rmv \mathbb{R}^{2\rmv\times\rmv 1} $ that denotes the mirror image of the $j$th \ac{pa} on these surfaces, where the index tuple $ (s\rmv,\rmv s') \in \mathcal{D} \triangleq \{(s\rmv,\rmv s') \in \mathcal{S}\rmv\rmv\times\rmv\rmv\mathcal{S} \} $ denotes the signal interaction order from the $ s'$th surface to the $ s$th surface before reaching the mobile agent. The two sets $ \mathcal{D}_{\mathrm{S}} \triangleq \{(s\rmv,\rmv s) \in \mathcal{S}\rmv\rmv\times\rmv\rmv\mathcal{S} \} $ with $ \vert \mathcal{D}_{\mathrm{S}}\vert = S $ and $ \mathcal{D}_{\mathrm{D}} \triangleq \{(s\rmv,\rmv s') \in \mathcal{S}\rmv\rmv\times\rmv\rmv\mathcal{S} | s\neq s'\} $ with $ \vert \mathcal{D}_{\mathrm{D}}\vert = S(S-1) $ contain the index tuples for single-bounce and double-bounce \acp{va}, respectively. Therefore, the maximum number of \acp{va} for \ac{pa} $j$ is given by $ \vert \mathcal{D}\vert = S+S(S-1) $ and $\mathcal{D} = \mathcal{D}_{\mathrm{S}} \cup \mathcal{D}_{\mathrm{D}}$. For each propagation path, its distance $d_{ss',n}^{(j)}$, \ac{aod} $\phi_{ss',n}^{(j)}$ and \ac{aoa} $\varphi_{ss',n}^{(j)}$ are modeled by $ d_{ss',n}^{(j)} = \|\V{p}_{n} - \V{p}_{ss',\mathrm{va}}^{(j)}\| $, $ \phi_{ss',n}^{(j)} = \angle (\V{p}_{\mathrm{\ip}ss',n}^{(j)}, \V{p}_{\mathrm{pa}}^{(j)}, \Delta\phi^{(j)})$ and $ \varphi_{ss',n}^{(j)} = \angle (\V{p}_{n}, \V{p}_{ss',\mathrm{va}}^{(j)}, \Delta\varphi_{n})$, and $\V{p}_{\mathrm{\ip}ss',n}^{(j)}$ denotes the \ac{ip} of the path on the first interacting surface $s'$. The geometric transformation between $\V{p}_{ss,\mathrm{va}}^{(j)}$, $\V{p}_{ss',\mathrm{va}}^{(j)}$ and $\V{p}_{\mathrm{\ip}ss',n}^{(j)}$, and $\V{p}_{s,\mathrm{\sfv}}$, $\V{p}_{s',\mathrm{\sfv}}$ and $\V{p}_{\mathrm{pa}}^{(j)}$ is provided in \cite[Section~IIA]{LeiVenTeaMey:TSP2023}.

The \ac{roi} is assumed to be partitioned into contiguous grid cells with index $i \in \mathcal{Q} \triangleq \{1,\dots,Q\} $, and $ Q $ is the total number of cells. As shown in Fig.~\ref{fig:FactorGraphSum}, each propagation path interacts with a subset of cells $ \mathcal{Q}_{ss',n}^{(j)} $, i.e., the traversed cells $ \tilde{\mathcal{Q}}_{ss',n}^{(j)} $ providing evidence of free space and the hit cells $ \tilde{\tilde{\mathcal{Q}}}_{ss',n}^{(j)} $ which denote path-object interactions and provide occupancy evidence. The remaining cells are denoted as $ \breve{\mathcal{Q}}_{ss',n}^{(j)} = \mathcal{Q} \backslash \{ \tilde{\mathcal{Q}}_{ss',n}^{(j)}  \cup \tilde{\tilde{\mathcal{Q}}}_{ss',n}^{(j)} \} $. The sets of cells are identified by performing ray casting \cite{Thrun_OGM2003,Kwon_TRob2019}.

%% file: InputFiles/SystemModel.tex
At each time $ n $, the state of the mobile agent is given by $ \RV{x}_{n} \triangleq [\RV{{\beta}}_{n}^{\mathrm{T}} \iist \rv{{\Delta\varphi}}_{n}]^{\mathrm{T}}$ with $ \RV{{\beta}}_{n}^{\mathrm{T}} \triangleq [\RV{p}_{n}^{\mathrm{T}} \iist \RV{v}_{n}^{\mathrm{T}}]^{\mathrm{T}} $. The state vector consists of the position $ \RV{p}_{n} $, the velocity $ \RV{v}_{n} =[\rv{v}_{\mathrm{x},n} \iist \rv{v}_{\mathrm{y},n}]^{\mathrm{T}} $ and the orientation of the azimuth array $ \rv{{\Delta\varphi}}_{n}$. The stacked agent state vector for all time steps up to $ n $ is given by $ \RV{x}_{1:n} \triangleq [\RV{x}_{1}^{\mathrm{T}} \ist\cdots\ist \RV{x}_{n}^{\mathrm{T}} ]^{\mathrm{T}} $. Following \cite{LeiVenTeaMey:TSP2023}, we account for the unknown and time-varying number of \acp{sfv} at each time step $n$ by introducing \acp{psfv} indexed by $s \in \{1,\dots, S_{n}\}$, where $ S_{n} $ represents the maximum possible number of \acp{psfv} that have produced at least one measurement so far. \Ac{psfv} states are modeled as $ \RV{y}_{s,n} \triangleq [\RV{p}_{s,\mathrm{\sfv}}^{\mathrm{T}} \iist \rv{{\rho}}_{s} \iist \rv{r}_{s,n}]^{\mathrm{T}}$ where $\RV{p}_{s,\mathrm{\sfv}}$ and $ \rv{{\rho}}_{s} $ denote the position and reflection coefficient for each ray interaction. The existence or non-existence of the $s$th \ac{psfv} is modeled by a binary random variable $\rv{r}_{s,n} \in \{1, 0\} $, and it exists if and only if $r_{s,n} = 1$.

The \acp{psfv} that have been detected either at a previous time $n'\rmv<\rmv n$, or at time $n$ but at a previous \ac{pa} $j'\rmv<\rmv j$ are referred to as \textit{legacy \acp{psfv}} with states $\underline{\RV{y}}_{s,n}^{(j)} \triangleq [ {\underline{\RV{p}}_{s,\mathrm{\sfv}}^{(j)\mathrm{T}}} \iist \underline{\rho}_{s}^{(j)} \iist  \underline{\rv{r}}_{s,n}^{(j)}]^{\mathrm{T}} $. The stacked \ac{psfv} state vector is given by $ \underline{\RV{y}}_{n}^{(1)} \rmv\rmv\triangleq\rmv\rmv [\underline{\RV{y}}_{1,n}^{(1)\mathrm{T}} \ist \cdots \ist  \underline{\RV{y}}_{S_{n-1}^{(J)} ,n}^{(1)\mathrm{T}} ]^{\mathrm{T}} $ at \ac{pa} $j\rmv=\rmv1$, or $ \underline{\RV{y}}_{n}^{(j)} \rmv\rmv\triangleq\rmv\rmv [\underline{\RV{y}}_{1,n}^{(j)\mathrm{T}} \ist \cdots \ist  \underline{\RV{y}}_{S_{n}^{(j-1)},n}^{(j)\mathrm{T}} ]^{\mathrm{T}} $ at \ac{pa} $j>1$. \textit{New \acp{sfv}} that generate measurements for the first time are modeled by a Poisson point process with mean $\mu_{\mathrm{n}}$ and conditional \ac{pdf} $f_{\mathrm{n}}( \overline{\V{p}}_{m,\mathrm{\sfv}}^{(j)}, \overline{\rho}_{m}^{(j)}| \V{x}_{n})$. The state of new \acp{psfv} is given by $ \overline{\RV{y}}_{m,n}^{(j)} \triangleq [\overline{\RV{p}}_{m,\mathrm{\sfv}}^{{(j)}\mathrm{T}} \iist  \overline{\rho}_{m}^{(j)} \iist  \overline{\rv{r}}_{m,n}^{(j)}]^{\mathrm{T}} $, $ m \in \{1,\dots, \rv{M}_n^{(j)}\} $. The state vector of all new \acp{psfv} at \ac{pa} $j$ is given by $ \overline{\RV{y}}_{n}^{(j)} \triangleq [\overline{\RV{y}}_{1,n}^{{(j)}\mathrm{T}} \ist\cdots\ist \overline{\RV{y}}_{\rv{M}_{n}^{(j)},n}^{{(j)}\mathrm{T}} ]^{\mathrm{T}} $. New \acp{psfv} become legacy \acp{psfv} after the next measurements update, either of the next \ac{pa} or at the next time they are observed. The number of legacy \acp{psfv} is updated according to $ S_{n}^{(1)} = S_{n-1}^{(J)} + M_{n}^{(1)} $ for $j=1$, and $ S_{n}^{(j)} = S_{n}^{(j-1)} + M_{n}^{(j)} $ for $j>1$. After the measurements of all $J$ \acp{pa} have been incorporated at time $n$, the state of all \ac{psfv} states at time $n$ is given by $ \RV{y}_{n} \triangleq [\underline{\RV{y}}_{n}^{{(J)}\mathrm{T}} \iist \overline{\RV{y}}_{n}^{{(J)}\mathrm{T}} ]^{\mathrm{T}} $, which is further stacked into the vector for all times up to $n$ given by $ \RV{y}_{1:n} \triangleq [\RV{y}_{1}^{\mathrm{T}} \ist\cdots\ist \RV{y}_{n}^{\mathrm{T}} ]^{\mathrm{T}} $. To account for the time-varying \ac{los} condition from \ac{pa} $ j $ to the agent, we define the \ac{pa} state as $ \underline{\RV{y}}_{0,n}^{(j)} \triangleq [\underline{\RV{p}}_{0,\mathrm{\sfv}}^{(j)\mathrm{T}} \iist \underline{\rv{{\rho}}}_{0}^{(j)} \iist \underline{\rv{r}}_{0,n}^{(j)}]^{\mathrm{T}}$, with $ \underline{\RV{p}}_{0,\mathrm{\sfv}}^{(j)} \triangleq \V{p}_{\mathrm{pa}}^{(j)} $ and $\underline{{\rho}}_{0}^{(j)} \triangleq 1 $. The set $\mathcal{D}_{n}^{(j)}$ represents all propagation paths at \ac{pa} $ j $ and time $ n $, and the set $\tilde{\mathcal{D}}_{n}^{(j)} \triangleq \{(0,0)\} \cup \mathcal{D}_{n}^{(j)}$ further includes the \ac{los}.

\subsection{Grid Cell State}
The occupancy state of each cell is modeled as a binary random variable $ \rv{o}_{i,n} \in \{0,1\}$ with $i \in \{1,\dots,Q\} $, where $ o_{i,n} = 0 $ indicates free state, $ o_{i,n} = 1 $ indicates occupied state. The corresponding probabilities are given by $ p(o_{i,n}=0) $ and $ p(o_{i,n}=1) = 1-p(o_{i,n}=0) $, respectively. The stacked vectors of all cell states are given by $\RV{o}_{n} \triangleq [\rv{o}_{1,n} \cdots \rv{o}_{Q,n} ]^{\mathrm{T}}$ at time $ n $, and $\RV{o}_{0:n} \triangleq [\RV{o}_{0}^{\mathrm{T}} \cdots \RV{o}_{n}^{\mathrm{T}} ]^{\mathrm{T}}$ for all times up to $ n $, respectively. Note that the grid cell states are physically time-invariant in a stationary environment. The time dependence is introduced solely to model the inference process. We define $\mathcal{V}_{ss',n}^{(j)}$ as the set of occupancy configurations that spatially support an existing path $ (s\rmv,\rmv s') $: $ \mathcal{V}_{ss',n}^{(j)} \rrmv\rrmv
\triangleq
\left\{ \V{o}_n \rmv\in \rmv\{0,1\}^{Q} | 
o_{i,n}\rmv=\rmv0,\ \rrmv \forall i\in \tilde{\mathcal{Q}}_{ss',n}^{(j)},\rmv
\; \exists i' \rmv\in \rmv\tilde{\tilde{\mathcal{Q}}}_{ss',n}^{(j)} \rrmv:\rmv o_{i',n}\rmv=\rmv1
\right\}$.

\vspace{-1mm}
\subsection{Measurement Model}
The mobile agent state and \ac{psfv} states relate to the distance measurements ${\rv{z}_\mathrm{d}}_{m,n}^{(j)}$, the \ac{aod} measurements $ {\rv{z}_\mathrm{\phi}}_{m,n}^{(j)} $, the \ac{aoa} measurements ${\rv{z}_\mathrm{\varphi}}_{m,n}^{(j)}$, and the normalized amplitude measurements ${\rv{z}_\mathrm{u}}_{m,n}^{(j)}\rmv\in\rmv [u_{\mathrm{de}}, \infty) $ via \acp{lhf}, which are assumed to be conditionally independent across $ m $. Note that the measurements are obtained by applying a parametric channel estimation algorithm to \ac{rf} measurements \cite{RichterPhD2005,HansenTSP2018,grebien2024SBL} in the pre-processing stage, where a detection threshold $ u_{\mathrm{de}} $ is used. The \acp{lhf} for the \ac{los} paths $ f_{\mathrm{P}}(\V{z}_{m,n}^{(j)}|\V{p}_{n}, u_{00,n}^{(j)}) $, for the single-bounce paths $ f_{\mathrm{SB}}(\V{z}_{m,n}^{(j)}|\V{p}_{n}, \V{p}_{s,\mathrm{\sfv}}^{(j)}, u_{ss,n}^{(j)}) $, for the double-bounce paths $ f_{\mathrm{DB}}(\V{z}_{m,n}^{(j)}|\V{p}_{n}, \V{p}_{s,\mathrm{\sfv}}^{(j)}, \V{p}_{s',\mathrm{\sfv}}^{(j)}, u_{ss',n}^{(j)}) $ and for the new paths $ f_{\mathrm{N}}(\V{z}_{m,n}^{(j)}|\V{p}_{n}, \overline{\V{p}}_{m,\mathrm{\sfv}}^{(j)}, \overline{u}_{m,n}^{(j)})  $, as well as the individual \acp{lhf} of the distance $ f_{ss'}^{(j)}({z_\mathrm{d}}_{m,n}^{(j)}) $, \ac{aod} $ f_{ss'}^{(j)}({z_\mathrm{\phi}}_{m,n}^{(j)}) $, \ac{aoa} $ f_{ss'}^{(j)}({z_\mathrm{\varphi}}_{m,n}^{(j)}) $ and the normalized amplitude $ f_{ss'}^{(j)}({z_\mathrm{u}}_{m,n}^{(j)}) $ are in line with those presented in \cite{MVASLAM_TWC2025Arxiv}.

The normalized amplitude $ u_{ss',n}^{(j)} $ is defined as the square root of the propagation path’s \ac{snr}. Assuming that the path amplitude follows free-space pathloss and is attenuated by the reflection coefficient $ \underline{{\rho}}_{s} $ after each surface reflection, $ u_{ss',n}^{(j)} $ is given by
\begin{align}
	{u}_{ss',n}^{(j)}
	= \beta_{ss',n}^{(j)} \sqrt{H_{\mathrm{rx}} P_{\mathrm{tx}}} \lambda_{\mathrm{c}}/ (4\pi d_{ss',n}^{(j)} \sigma_{\mathrm{n}})
	\label{eq:friis}
\end{align}
where $ \lambda_{\mathrm{c}} $ is the wavelength at center frequency, $ P_{\mathrm{tx}} $ is the known total transmit power, $ \sigma_{\mathrm{n}} $ is the noise \ac{std} provided by the parametric channel estimator, and $ \beta_{ss',n}^{(j)} $ is the amplitude attenuation factor given as $ \beta_{ss',n}^{(j)} = \underline{{\rho}}_{s} \underline{{\rho}}_{s'}$ for $s\neq s'$, and $ \beta_{ss,n}^{(j)} = \underline{{\rho}}_{s} $. The normalized amplitude for new paths $ \overline{u}_{m,n} $ is also obtained as in \eqref{eq:friis} with $ \beta_{m,n}^{(j)} = \overline{\rho}_{m} $. The normalized amplitude is directly related to the detection probability $ p_{\mathrm{d},ss'}^{(j)} = p_{\mathrm{d}}({u}_{ss',n}^{(j)}) $ with $(s,s') \in \tilde{\mathcal{D}}_{n}^{(j)} $ and measurement variances via Fisher information \cite{MVASLAM_TWC2025Arxiv}.

\begin{figure*}[!t]	
	\begin{align}
		& f(\V{x}_{0:n},\V{y}_{0:n},\V{o}_{0:n},\underline{\V{a}}_{1:n},\bar{\V{a}}_{1:n}\mid	\V{z}_{1:n}) \nn \\
		&\propto
		f(x_0)\prod_{s=1}^{S_0} f(\V{y}_{s,0})\prod_{i=1}^{Q} f(o_{i,0})
		\left( \prod_{n'=1}^{n} f(\V{x}_{n'} \mid \V{x}_{n'-1}) f(\V{o}_{n'} \mid \V{o}_{n'-1}) \right)
		\prod_{s'=1}^{S_{n'}}
		f(\underline{\V{y}}_{s',n'} \mid \V{y}_{s',n'-1})
		\left( \prod_{j'=2}^{J} \prod_{s'=1}^{S_{n'}^{(j')}} f(\underline{\V{y}}_{s',n'}^{(j')} \mid \V{y}_{s',n'}^{(j'-1)}) \hspace{-0.5mm} \right) 
		\nn \\[-0.3em]
		&\hspace{0.5mm} \times \hspace{-1mm}
		\left(
		\prod_{j=1}^{J}
		\underline{q}_{\mathrm{P}}(\V{x}_{n'}, \underline{\V{y}}_{0,n'}^{(j)}, \V{o}_{n'}, \underline{a}_{00,n'}^{(j)};\V{z}_{n'}^{(j)}) \prod_{m'=1}^{M_{n'}^{(j)}}
		\psi( \underline{a}_{00,n'}^{(j)}, \bar{a}_{m',n'}^{(j)})  \hspace{-0.5mm}
		\right)\hspace{-1mm}
		\prod_{j=1}^{J} \hspace{-1mm}
		\left(
		\prod_{s=1}^{S_{n'}^{(j)}}
		\underline{q}_{\mathrm{S}}(\V{x}_{n'}, \underline{\V{y}}_{s,n'}^{(j)}, \V{o}_{n'}, \underline{a}_{ss,n'}^{(j)};\V{z}_{n'}^{(j)}) \prod_{m'=1}^{M_{n'}^{(j)}}
		\psi( \underline{a}_{ss,n'}^{(j)}, \bar{a}_{m',n'}^{(j)} ) \hspace{-0.5mm}
		\right)
		\nn \\[-0.25em]
		&\hspace{0.5mm} \times \hspace{-1mm}
		\left(
		\prod_{s'=1,s'\neq s}^{S_{n'}^{(j)}}
		\underline{q}_{\mathrm{D}}(\V{x}_{n'}, \underline{\V{y}}_{s,n'}^{(j)},\underline{\V{y}}_{s',n'}^{(j)}, \V{o}_{n'}, \underline{a}_{ss',n'}^{(j)};  \V{z}_{n'}^{(j)})
		\right)\hspace{-2mm}
		\left(
		\prod_{m'=1}^{M_{n'}^{(j)}}
		\psi( \underline{a}_{ss',n'}^{(j)}, \bar{a}_{m',n'}^{(j)} )
		\prod_{m=1}^{M_{n'}^{(j)}}
		\overline{q}_{\mathrm{N}}( \V{x}_{n'}, \overline{\V{y}}_{m,n'}^{(j)}, \overline{a}_{m,n'}^{(j)};\V{z}_{n'}^{(j)}) \hspace{-0.5mm} \right)\hspace{-0mm} \, .
		\label{eq:JointPosteriorPDF}\\[-9.5mm]\nn
	\end{align}	
	\hrulefill
	\vspace*{-4mm}
\end{figure*}
\subsection{State Evolution}
The agent state, legacy \ac{psfv} states and cell states are assumed to evolve independently across $ j $ and $ n $ according to state-transition \acp{pdf} $f(\V{x}_{n}|\V{x}_{n-1})$, $f(\underline{\V{y}}_{s,n}|\V{y}_{s,n-1})$ and $f\left(o_{i,n} \mid o_{i,n-1}\right) $. The state-transition \ac{pdf} for grid cells is modeled as $f\left(o_{i,n} \mid o_{i,n-1}\right) = \delta(o_{i,n}-o_{i,n-1}) $ for stationary environments. In line with \cite{Florian_TSP2017,Florian_Proceeding2018, Erik_SLAM_TWC2019, LeiVenTeaMey:TSP2023}, the state-transition \ac{pdf} $ f(\underline{\V{y}}_{s,n}|\V{y}_{s,n-1}) = f(\underline{\V{p}}_{s,\mathrm{\sfv}}, \underline{\rho}_{s}, \underline{r}_{s,n} | \V{p}_{s,\mathrm{\sfv}}, {\rho}_{s}, r_{s,n-1}) $ at \ac{pa} $j=1$ is defined as
\begin{align}
	\hspace*{-3mm}  f(\underline{\V{p}}_{s,\mathrm{\sfv}}, \underline{\rho}_{s}, \underline{r}_{s,n} | \V{p}_{s,\mathrm{\sfv}}, {\rho}_{s}, 1) & \nn \\
	&\hspace*{-20mm} = \begin{cases}
		(1-p_{\mathrm{s}})f_{\mathrm{D}}(\underline{\V{p}}_{s,\mathrm{\sfv}}, \underline{\rho}_{s}), 	&\underline{r}_{s,n}= 0\\
		p_{\mathrm{s}} f(\underline{\V{p}}_{s,\mathrm{\sfv}}, \underline{\rho}_{s} | \V{p}_{s,\mathrm{\sfv}}, {\rho}_{s}), 											&\underline{r}_{s,n}= 1  
	\end{cases}
	\label{eq:STpdf_SFV1} 
\end{align}
for $r_{s,n-1} = 1$. For a nonexistent \ac{sfv} at the previous time, i.e., $r_{s,n-1} = 0$, it is defined as
\begin{align}
	\hspace*{-2.5mm} f(\underline{\V{p}}_{s,\mathrm{\sfv}}, \underline{\rho}_{s}, \underline{r}_{s,n} | \V{p}_{s,\mathrm{\sfv}}, {\rho}_{s}, 0) \rrmv = \rrmv 
	\begin{cases}
		f_{\mathrm{D}}(\underline{\V{p}}_{s,\mathrm{\sfv}}, \underline{\rho}_{s}), 	& \hspace*{-2.1mm} \underline{r}_{s,n}\rmv = \rmv 0\\
		0, 											& \hspace*{-2.1mm} \underline{r}_{s,n}\rmv = \rmv 1 \, .
	\end{cases}
	\label{eq:STpdf_SFV2} 
\end{align}
To account for sequential updating, we further define $ f(\underline{\V{p}}_{s,\mathrm{\sfv}}^{(j)}, \underline{\rho}_{s}^{(j)}, \underline{r}_{s,n}^{(j)} | \V{p}_{s,\mathrm{\sfv}}^{(j-1)}, {\rho}_{s}^{(j-1)}, r_{s,n}^{(j-1)}) $ for $j\geq2$, reads
\begin{align}
	\hspace*{-0mm}  & f(\underline{\V{p}}_{s,\mathrm{\sfv}}^{(j)}, \underline{\rho}_{s}^{(j)}, \underline{r}_{s,n}^{(j)} | \V{p}_{s,\mathrm{\sfv}}^{(j-1)}, {\rho}_{s}^{(j-1)}, 1 )  \nn \\ 
	&\hspace*{14mm} = 
	\begin{cases}
		f_{\mathrm{D}}( \underline{\V{p}}_{s,\mathrm{\sfv}}^{(j)}, \underline{\rho}_{s}^{(j)} ), 	& \hspace*{-0.7mm} \underline{r}_{s,n}^{(j)}= 0\\
		f(\underline{\V{p}}_{s,\mathrm{\sfv}}^{(j)}, \underline{\rho}_{s}^{(j)}|\V{p}_{s,\mathrm{\sfv}}^{(j-1)}, \underline{\rho}_{s}^{(j-1)} ), 											& \hspace*{-0.7mm} \underline{r}_{s,n}^{(j)}= 1 \, ,
	\end{cases}
	\label{eq:STpdf_SFV3}  \vspace*{0mm}\\
	\hspace*{-15mm} & f(\underline{\V{p}}_{s,\mathrm{\sfv}}^{(j)}, \underline{\rho}_{s}^{(j)}, \underline{r}_{s,n}^{(j)} | \V{p}_{s,\mathrm{\sfv}}^{(j-1)}, {\rho}_{s}^{(j-1)}, 0 )  \nn \\ 
	&\hspace*{14mm} = 
	\begin{cases}
		f_{\mathrm{D}}( \underline{\V{p}}_{s,\mathrm{\sfv}}^{(j)}, \underline{\rho}_{s}^{(j)} ), 	& \hspace*{19mm} \underline{r}_{s,n}^{(j)}= 0\\
		0, 											& \hspace*{19mm} \underline{r}_{s,n}^{(j)}= 1 \, .
	\end{cases}
	\label{eq:STpdf_SFV4} 
\end{align}
The \ac{pdf} $ f_{\mathrm{D}}(\cdot) $ is an arbitrary ``dummy \ac{pdf}'' to complete the statistics. The initial prior \acp{pdf} $f(\V{x}_{0})$, $f(\V{y}_{s,0})$ for $ s \in \mathcal{S}_{0} $ and $f(\V{o}_{0})$ at time $n=0$ are assumed to be known. Note that the initial prior \ac{pdf} $f(\V{o}_{0})$ can be modeled as uniform when no prior knowledge of the \ac{roi} is available, or can be obtained from an established map generated by other agents or sensors.

\vspace{-1mm}
\subsection{Data Association}
\Ac{da} between measurements and \acp{psfv} is represented by two complementary \ac{da} vectors: the \emph{\ac{psfv}-oriented} association vector $ \RV{\underline{a}}_{n}^{(j)} \triangleq [\rv{\underline{a}}_{00,n}^{(j)} \ist \rv{\underline{a}}_{11,n}^{(j)} \ist \cdots \ist  \rv{\underline{a}}_{\rv{S}_{n}^{(j)}\rmv\rv{S}_{n}^{(j)},n}^{(j)}]^{\mathrm{T}} $ and the \emph{measurement-oriented} association vector $ \RV{\overline{a}}_{n}^{(j)} \triangleq [\rv{\overline{a}}_{1,n}^{(j)} \ist \cdots \ist \rv{\overline{a}}_{\rv{M}_{n}^{(j)},n}^{(j)}]^{\mathrm{T}} $, see \cite{ Florian_Proceeding2018, LeiVenTeaMey:TSP2023, MVASLAM_TWC2025Arxiv} for further details.

\subsection{Factor Graph}
Using Bayes' rule and independence assumptions related to the state-transition \acp{pdf}, the prior \acp{pdf}, and the \ac{lhf}, the joint posterior \ac{pdf} $f(\V{x}_{0:n},\V{y}_{0:n},\V{o}_{0:n},\underline{\V{a}}_{1:n},\bar{\V{a}}_{1:n}\mid z_{1:n})$ is given by \eqref{eq:JointPosteriorPDF}, where the individual \ac{lhf} related factors are introduced as follows. 

\subsubsection*{For the \ac{los} path related to \ac{pa} $j$} the pseudo \ac{lhf} $\underline{q}_{\mathrm{P}}(\V{x}_{n}, \underline{\V{y}}_{0,n}^{(j)}, \V{o}_{n}, \underline{a}_{00,n}^{(j)};\V{z}_{n}^{(j)}) = \underline{q}_{\mathrm{P}}(\V{x}_{n}, \underline{r}_{0,n}^{(j)}, \V{o}_{n}, \underline{a}_{00,n}^{(j)};\V{z}_{n}^{(j)})$ is given by
\vspace{-1mm}
\begin{align}
	& \hspace*{0mm}\underline{q}_{\mathrm{P}}(\V{x}_{n},  \underline{r}_{0,n}^{(j)}=1, \V{o}_{n},  \underline{a}_{00,n}^{(j)};\V{z}_{n}^{(j)}) \nn \\[0.5mm]
	& \hspace*{8mm} \triangleq
	\begin{cases}
		\dfrac{ f_{\mathrm{P}}(\V{z}_{m,n}^{(j)}|\V{p}_{n}, u_{00,n}^{(j)})  p_{\mathrm{d},00}^{(j)}  } {\mu_{\mathrm{fa}} f_{\mathrm{fa}}(\V{z}_{m,n}^{(j)})}, 										& \underline{a}_{00,n}^{(j)} \in \mathcal{M}_{n}^{(j)} \\[4mm]
		1 - p_{\mathrm{d},00}^{(j)} ,										& \underline{a}_{00,n}^{(j)} = 0.
	\end{cases}
	\label{eq:LHFPA}\\[-7mm]\nn
\end{align}
for the case $ \V{o}_{n} \in \mathcal{V}_{0,n}^{(j)}$, otherwise $ \underline{q}_{\mathrm{P}}(\V{x}_{n}, \underline{r}_{0,n}^{(j)}=0, \V{o}_{n}, \underline{a}_{00,n}^{(j)};\V{z}_{n}^{(j)}) \triangleq \delta(\underline{a}_{00,n}^{(j)}) $ for $\V{o}_{n} \in \{0,1\}^{Q}$.

\begin{figure*}[t]
	\centering
	\vspace*{0mm}
	\hspace*{-4mm}\subfloat[\label{subfig:FactorGraphAllall}]
	{\hspace*{2mm}\includegraphics[scale=0.98]{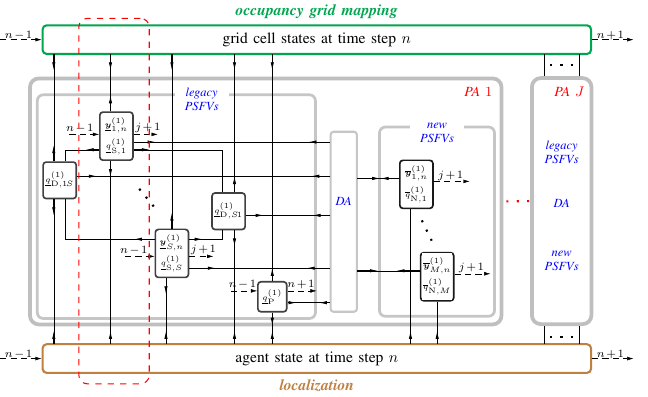}}
	\hspace*{0mm}\subfloat[\label{subfig:FactorGraphSS}]
	{\hspace*{1mm}\includegraphics[scale=0.97]{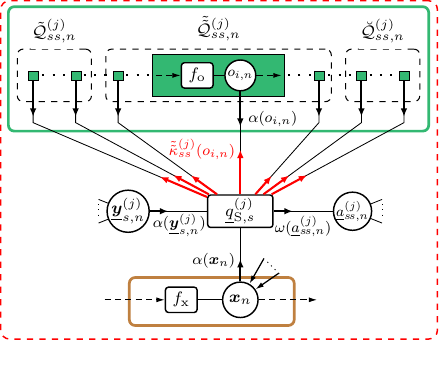}}
	\caption{Factor graph representation of the joint posterior \ac{pdf} \eqref{eq:JointPosteriorPDF}. (a) shows the message propagation between the subgraphs, as well as iterations over time and \acp{pa}. The black square and circle symbols denote factor nodes and variable nodes, respectively. Dashed arrows indicate messages propagated forward in time. As an example, (b) shows the variable nodes, factor nodes, and messages for a single-bounce path. The following short notations are used: $S \triangleq S_{n}^{(j)} $, $M \triangleq M_{n}^{(j)} $, $f_{\mathrm{x}} \triangleq  f(\V{x}_{n}|\V{x}_{n-1}) $, $f_{\mathrm{o}} \triangleq  f(o_{i,n}|o_{i,n-1}) $, $ \underline{q}_{\mathrm{P}}^{(j)} $, $ \underline{q}_{\mathrm{S},s}^{(j)} $, $ \underline{q}_{\mathrm{D},\rmv s\rmv s'}^{(j)} $ and $ \overline{q}_{\mathrm{N},m}^{(j)} $ represent the pseudo \acp{lhf} of the \ac{los} paths \eqref{eq:LHFPA}, single-bounce paths \eqref{eq:LHFSpath}, double-bounce paths \eqref{eq:LHFDpath} and new paths \eqref{eq:LHFNew}, respectively. }
	\label{fig:FactorGraphAll}
\end{figure*}

\subsubsection*{For single-bounce paths related to PA $j$} $(s\rmv,\rmv s) \in \mathcal{D}_{\mathrm{S},n}^{(j)}$, the pseudo \ac{lhf} $\underline{q}_{\mathrm{S}}(\V{x}_{n}, \underline{\V{y}}_{s,n}^{(j)}, \V{o}_{n}, \underline{a}_{ss,n}^{(j)};\V{z}_{n}^{(j)}) = \underline{q}_{\mathrm{S}}(\V{x}_{n}, \underline{\V{p}}_{s,\mathrm{\sfv}}^{(j)}, \underline{r}_{s,n}^{(j)}, \underline{\rho}_{s}^{(j)},\V{o}_{n}, \underline{a}_{ss,n}^{(j)};\V{z}_{n}^{(j)})$ is given by 
\begin{align}
	& \hspace*{0mm} \underline{q}_{\mathrm{S}}(\V{x}_{n}, \underline{\V{p}}_{s,\mathrm{\sfv}}^{(j)}, \underline{r}_{s,n}^{(j)} = 1, \underline{\rho}_{s}^{(j)}, \V{o}_{n}, \underline{a}_{ss,n}^{(j)};\V{z}_{n}^{(j)}) \nonumber \\[1mm]
	& \hspace*{3mm} \triangleq 
	\begin{cases}
		\dfrac{ f_{\mathrm{SB}}(\V{z}_{m,n}^{(j)}|\V{p}_{n}, \V{p}_{s,\mathrm{\sfv}}^{(j)}, u_{ss,n}^{(j)}) p_{\mathrm{d},ss}^{(j)} } {\mu_{\mathrm{fa}} f_{\mathrm{fa}}(\V{z}_{m,n}^{(j)})}, 								& \hspace*{-2mm} \underline{a}_{ss,n}^{(j)} \in \mathcal{M}_{n}^{(j)} \\[4mm]
		1 - p_{\mathrm{d},ss}^{(j)},						& \hspace*{-2mm} \underline{a}_{ss,n}^{(j)} = 0
	\end{cases}
	\hspace*{-2mm}\label{eq:LHFSpath}\\[-7mm]\nn
\end{align}
for the case $ \V{o}_{n} \in \mathcal{V}_{ss,n}^{(j)}$, otherwise $\underline{q}_{\mathrm{S}}(\V{x}_{n}, \underline{\V{p}}_{s,\mathrm{\sfv}}^{(j)}, \underline{r}_{s,n}^{(j)} = 0, \underline{\rho}_{s}^{(j)}, \V{o}_{n}, \underline{a}_{ss,n}^{(j)}; \V{z}_{n}^{(j)})  \triangleq \delta(\underline{a}_{ss,n}^{(j)})$ for $\V{o}_{n} \in \{0,1\}^{Q}$.

\subsubsection*{For double-bounce paths related to $(s\rmv,\rmv s') \in \mathcal{D}_{\mathrm{D},n}^{(j)}$} the pseudo \ac{lhf} $\underline{q}_{\mathrm{D}}(\V{x}_{n}, \underline{\V{y}}_{s,n}^{(j)},\underline{\V{y}}_{s',n}^{(j)}, \V{o}_{n}, \underline{a}_{ss',n}^{(j)};  \V{z}_{n}^{(j)}) =  \underline{q}_{\mathrm{D}}(\V{x}_{n}, \underline{\V{p}}_{s,\mathrm{\sfv}}^{(j)},  \underline{r}_{s,n}^{(j)}, \underline{\rho}_{s}^{(j)}, \underline{\V{p}}_{s',\mathrm{\sfv}}^{(j)}, \underline{r}_{s',n}^{(j)}, \underline{\rho}_{s'}^{(j)}, \V{o}_{n}, \underline{a}_{ss',n}^{(j)};\V{z}_{n}^{(j)}) $ is given by
\begin{align}
	& \hspace*{-4mm} \underline{q}_{\mathrm{D}}(\V{x}_{n}, \underline{\V{p}}_{s,\mathrm{\sfv}}^{(j)}, 1, \underline{\rho}_{s}^{(j)}, \underline{\V{p}}_{s',\mathrm{\sfv}}^{(j)}, 1, \underline{\rho}_{s'}^{(j)}, \V{o}_{n}, \underline{a}_{ss',n}^{(j)};\V{z}_{n}^{(j)}) \nonumber \\[1mm]
	& \hspace*{-2mm} \triangleq 
	\begin{cases}
		f_{\mathrm{DB}}(\V{z}_{m,n}^{(j)}|\V{p}_{n}, \V{p}_{s,\mathrm{\sfv}}^{(j)}, \V{p}_{s',\mathrm{\sfv}}^{(j)}, u_{ss',n}^{(j)}) \\ 
		\hspace*{0mm} \hspace{0.3em} \times \hspace{0.3em} 
		\dfrac{p_{\mathrm{d},ss'}^{(j)} } {\mu_{\mathrm{fa}}  f_{\mathrm{fa}}(\V{z}_{m,n}^{(j)})}, 										& \hspace*{-6mm} \underline{a}_{ss',n}^{(j)} \in \mathcal{M}_{n}^{(j)} \\[4mm]
		1 - p_{\mathrm{d},ss'}^{(j)} ,										& \hspace*{-6mm} \underline{a}_{ss',n}^{(j)} = 0
	\end{cases}
	\label{eq:LHFDpath}\\[-7mm]\nn 
\end{align}
for the case $ \V{o}_{n} \in \mathcal{V}_{ss',n}^{(j)}$, otherwise $ \underline{q}_{\mathrm{D}}(\cdots) \triangleq \delta(\underline{a}_{ss',n}^{(j)})$ for $\V{o}_{n} \in \{0,1\}^{Q}$.

\subsubsection*{For new paths} we assume that the new path is independent of the grid cell states, the pseudo \ac{lhf} $ \overline{q}_{\mathrm{N}}( \V{x}_{n}, \overline{\V{y}}_{m,n}^{(j)}, \overline{a}_{m,n}^{(j)};\V{z}_{n}^{(j)}) = \overline{q}_{\mathrm{N}}(\V{x}_{n}, \overline{\V{p}}_{m,\mathrm{\sfv}}^{(j)},  \overline{r}_{m,n}^{(j)}=1, \overline{\rho}_{m}^{(j)}, \\ \overline{a}_{m,n}^{(j)};\V{z}_{n}^{(j)}) $ is given 
\vspace{-1mm}
\begin{align}
	& \hspace*{0mm} \overline{q}_{\mathrm{N}}(\V{x}_{n}, \overline{\V{p}}_{m,\mathrm{\sfv}}^{(j)},  \overline{r}_{m,n}^{(j)} =1, \overline{\rho}_{m}^{(j)}, \overline{a}_{m,n}^{(j)};\V{z}_{n}^{(j)}) \nonumber \\[1mm]
	& \hspace*{5mm} \triangleq 
	\begin{cases}
		0,										&\hspace*{0mm} \overline{a}_{m,n}^{(j)} \in \tilde{\mathcal{D}}_{n}^{(j)} \\[2mm]
		f_{\mathrm{N}}(\V{z}_{m,n}^{(j)}|\V{p}_{n}, \overline{\V{p}}_{m,\mathrm{\sfv}}^{(j)}, \overline{u}_{m,n}^{(j)})   \\
		\hspace*{0mm} \hspace{0.3em} \times \hspace{0.3em} 
		\dfrac{ \mu_{\mathrm{n}} f_{\mathrm{n}}(\overline{\V{p}}_{m,\mathrm{\sfv}}^{(j)}, \overline{\rho}_{m}^{(j)} | \V{x}_{n})} {\mu_{\mathrm{fa}} f_{\mathrm{fa}}(\V{z}_{m,n}^{(j)}) } 
		,  										&\hspace*{0mm} \overline{a}_{m,n}^{(j)} = 0
	\end{cases}
	\label{eq:LHFNew}\\[-7mm]\nn 
\end{align}
and $ \overline{q}_{\mathrm{N}}(\V{x}_{n}, \overline{\V{p}}_{m,\mathrm{\sfv}}^{(j)}, 0, \overline{\rho}_{m}^{(j)}, \overline{a}_{m,n}^{(j)};\V{z}_{n}^{(j)}) \triangleq f_{\mathrm{D}}(\overline{\V{p}}_{m,\mathrm{\sfv}}^{(j)}, \overline{\rho}_{m}^{(j)}) $.

\subsection{Detection and State Estimation}
Using all measurements up to time $ n $, we sequentially detect \acp{psfv} based on the marginal posterior existence probabilities $f(r_{s,n}=1 | \V{z}_{1:n}) = \int f( \V{p}_{s,\mathrm{\sfv}}, \rho_{s}, r_{s,n}=1 |\V{z}_{1:n}) \mathrm{d} \V{p}_{s,\mathrm{\sfv}} \mathrm{d} \rho_{s} $. A \ac{psfv} is declared to exist (i.e., detected) if $ f( r_{s,n}=1 | \V{z}_{1:n}) > p_{\mathrm{\sfv}}$, where $p_{\mathrm{\sfv}}$ is the existence probability threshold. Since introducing new \acp{psfv} causes the number of \acp{psfv} to grow with time $ n $, \acp{psfv} with existence probabilities below a threshold $ p_{\mathrm{pr}} $ are removed from the state space (``pruned''). We determine the \ac{mmse} estimate by calculating the marginal posterior \acp{pdf} $f( \V{p}_{s,\mathrm{\sfv}} |r_{s,n}=1,\V{z}_{1:n}) = (\int f( \V{p}_{s,\mathrm{\sfv}}, \rho_{s}, r_{s,n}=1 |\V{z}_{1:n}) \mathrm{d} \rho_{s}) / f( r_{s,n}=1 | \V{z}_{1:n}) $, $ f(  \rho_{s} | r_{s,n}=1, \V{z}_{1:n}) $ and $ f(  o_{i,n} | \V{z}_{1:n}) $. The marginal posterior \acp{pdf} cannot be obtained analytically by direct marginalization of \eqref{eq:JointPosteriorPDF}, therefore we perform sequential message-passing by means of \ac{spa} on a factor graph to obtain approximations (beliefs) of these marginal posterior \acp{pdf}, for instance, $ \tilde{f}\left(o_{i,n}\right) $ is obtained as an approximation of $ f( o_{i,n} | \V{z}_{1:n}) $.

\section{The Proposed SPA Method}
\vspace{-0.5mm}
Due to space limitations, the full derivation of all \ac{spa} messages is omitted, and only the key messages relevant to the cell states are presented.

\subsection{Selected SPA Messages}
For all legacy \acp{psfv} $ s\in \mathcal{S}_{n-1} $, the predicted messages $\alpha(\underline{\V{y}}_{s,n}) = \alpha(\underline{\V{p}}_{s,\mathrm{\sfv}}, \underline{r}_{s,n}, \underline{\rho}_{s}) $ are given by
\vspace{-1mm}
\begin{align}
	&\alpha(\underline{\V{p}}_{s,\mathrm{\sfv}}, \underline{r}_{s,n}, \underline{\rho}_{s} )  \nn \\[-1.5mm] & \hspace*{5mm} = \sum_{r_{s,n-1} \in \{0,1\}} \int \rmv f(\underline{\V{p}}_{s,\mathrm{\sfv}}, \underline{r}_{s,n}, \underline{\rho}_{s} | \V{p}_{s,\mathrm{\sfv}}, r_{s,n-1}, \rho_{s}) \nn  \\[-0.5mm] & \hspace*{13mm} \times \tilde{f}(\V{p}_{s,\mathrm{\sfv}}, r_{s,n-1}, \rho_{s}) \mathrm{d} \V{p}_{s,\mathrm{\sfv}} \mathrm{d}\rho_{s}  \, .	
	\label{eq:prePSFV}
\end{align}
For all grid cells, the predicted messages are given by 	
\begin{align}
	\alpha(o_{i,n})
	= \sum_{o_{i,n-1}}
	f\left(o_{i,n} \mid o_{i,n-1}\right)\,
	\tilde{f}\left(o_{i,n-1}\right) \, .
	\label{eq:preGrid} \\[-7mm]\nn
\end{align}
Accordingly, we further define the predicted path-validity message $ \alpha_{\mathcal{V}_{ss',n}^{(j)}} $, given by
\vspace{-1mm}
\begin{align}
	\alpha_{\mathcal{V}_{ss',n}^{(j)}} \rrmv = \hspace*{-0mm} \Big( \rmv\rrmv \prod_{i\in \tilde{\mathcal{Q}}_{ss',n}^{(j)}}\hspace*{-2mm} \alpha(o_{i,n}=0)  \Big) \Big( 1-\hspace*{-4mm}\prod_{i'\in \tilde{\tilde{\mathcal{Q}}}_{ss',n}^{(j)}} \hspace*{-3mm} (1-\alpha(o_{i',n}=1))  \Big) \, .
	\label{eq:predOccupancyPMF}  \\[-8.5mm]\nn
\end{align} 
Note that in \eqref{eq:predOccupancyPMF}, cells in $\breve{\mathcal{Q}}_{ss',n}^{(j)}$ are unconstrained and marginalize to one over $o_{i,n} \rmv\in \rmv\{0,1\}$, so their messages are omitted. For the traversed set $ \tilde{\mathcal{Q}}_{ss',n}^{(j)} $, occupied cells along the path are penalized, since they may correspond to obstacles that block the propagation path. Conversely, for the hit set $\tilde{\tilde{\mathcal{Q}}}_{ss',n}^{(j)}$, the model aggregates occupancy evidence over the candidate interaction region, thereby reducing the influence of individual free cells on the path validity evaluation. Thus, \eqref{eq:predOccupancyPMF} penalizes potential blockages along the path while avoiding overly strict rejection due to isolated free cells near the uncertain reflection location. Overall, the model leverages prior occupancy-map information to evaluate path validity while maintaining robustness to local occupancy uncertainty.
\vspace{-0.5mm}

\subsubsection{Measurement Evaluation for \ac{los} Paths}
The messages $\omega(\underline{a}_{00,n}^{(j)})$ propagate from the $j$th \ac{pa} related factor node $\underline{q}_{\mathrm{P}}(\V{x}_{n}, \underline{\V{y}}_{0,n}^{(j)}, \V{o}_{n}, \underline{a}_{00,n}^{(j)}; \V{z}_{n}^{(j)}) $ to the feature-oriented association variable nodes $\underline{a}_{00,n}^{(j)}$ are given by
\begin{align}
	& \omega(\underline{a}_{00,n}^{(j)}) = \rrmv \underbracea{ \int \alpha(\V{x}_{n}) \alpha_{\mathcal{V}_{00,n}^{(j)}} \underline{q}_{\mathrm{P}}(\V{x}_{n}, \underline{r}_{0,n}^{(j)}=1, \V{o}_{n}, \underline{a}_{00,n}^{(j)};\V{z}_{n}^{(j)}) }_{\hspace*{20mm} R_{00,n}^{(j)}} \nn \\ & \times \underbraced{ \alpha(\underline{\V{p}}_{0,\mathrm{\sfv}}^{(j)}, \underline{r}_{0,n}^{(j)}=1, \underline{\rho}_{0}^{(j)}) 
		\mathrm{d} \V{x}_{n} }  + \underbrace{ 1(\underline{a}_{00,n}^{(j)}) \alpha_{0,n}^{(j)} }_{\hspace*{0mm} \overline{R}_{00,n}^{(j)}} 
	\label{eq:MeaEvaLos}
\end{align}
where $\alpha(\V{x}_{n})$ denotes the agent predicted message, and $\alpha_{s,n}^{(j)}$ is related to the nonexistence probability of \ac{psfv} $s$.

\subsubsection{Measurement Evaluation for Legacy \acp{psfv}}
\paragraph{Single-Bounce Propagation Path} For $s=s'$, the messages $\omega(\underline{a}_{ss,n}^{(j)})$ propagate from the single-bounce \ac{pva} related factor node $\underline{q}_{\mathrm{S}}(\V{x}_{n}, \underline{\V{y}}_{s,n}^{(j)}, \V{o}_{n}, \underline{a}_{ss,n}^{(j)};\V{z}_{n}^{(j)}) $ to the association variable nodes $\underline{a}_{ss,n}^{(j)}$ and are given by
\begin{align}
	&  \omega(\underline{a}_{ss,n}^{(j)}) = \rrmv \underbracea{ \int\rrmv\rrmv\rmv\int\rrmv\rrmv\rmv\int \alpha(\V{x}_{n}) \alpha_{\mathcal{V}_{ss,n}^{(j)}} \alpha(\underline{\V{p}}_{s,\mathrm{\sfv}}^{(j)}, \underline{r}_{s,n}^{(j)}=1, \underline{\rho}_{s}^{(j)} )  }_{\hspace*{20mm} R_{ss,n}^{(j)}} 
	\nn \\ & 
	\underbraced{ \underline{q}_{\mathrm{S}}(\V{x}_{n}, \underline{\V{p}}_{s,\mathrm{\sfv}}^{(j)}, \underline{r}_{s,n}^{(j)}=1, \underline{\rho}_{s}^{(j)}, \V{o}_{n}, \underline{a}_{ss,n}^{(j)};\V{z}_{n}^{(j)})   
		\mathrm{d} \V{x}_{n} \mathrm{d} \underline{\V{p}}_{s,\mathrm{\sfv}}^{(j)} \mathrm{d} \underline{\rho}_{s}^{(j)} } 
	\nn \\ &
	+ \underbrace{ 1(\underline{a}_{ss,n}^{(j)}) \alpha_{s,n}^{(j)} }_{\hspace*{0mm} \overline{R}_{ss,n}^{(j)}}\, .
	\label{eq:MeaEvaPSFV1}
\end{align}

\paragraph{Double-Bounce Propagation Path} For $s\neq s'$, the messages $\omega(\underline{a}_{ss',n}^{(j)})$ propagate from the double-bounce \ac{pva} related factor node $\underline{q}_{\mathrm{D}}(\V{x}_{n}, \underline{\V{y}}_{s,n}^{(j)},\underline{\V{y}}_{s',n}^{(j)}, \V{o}_{n}, \underline{a}_{ss',n}^{(j)};  \V{z}_{n}^{(j)}) $ to the feature-oriented association variable nodes $\underline{a}_{ss',n}^{(j)}$ given by
\begin{align}
	& \omega(\underline{a}_{ss',n}^{(j)}) = \rrmv \underbraceaa{ \int \rrmv\cdots \rrmv\rmv\int \alpha(\V{x}_{n}) \alpha_{\mathcal{V}_{ss',n}^{(j)}} \alpha(\underline{\V{p}}_{s,\mathrm{\sfv}}^{(j)}, \underline{r}_{s,n}^{(j)} \rrmv=\rrmv 1, \underline{\rho}_{s}^{(j)} ) }
	\nn \\ & 
	\underbraceb{ \times \alpha(\underline{\V{p}}_{s',\mathrm{\sfv}}^{(j)}, \underline{r}_{s',n}^{(j)} \rrmv=\rrmv 1, \underline{\rho}_{s'}^{(j)} )  \underline{q}_{\mathrm{D}}(\V{x}_{n}, \underline{\V{p}}_{s,\mathrm{\sfv}}^{(j)},  \underline{r}_{s,n}^{(j)}\rrmv=\rrmv 1, \underline{\rho}_{s}^{(j)}, \underline{\V{p}}_{s',\mathrm{\sfv}}^{(j)}, }_{\hspace*{4mm} R_{ss',n}^{(j)}}  
	\nn \\ & 
	\underbraced { \underline{r}_{s',n}^{(j)}=1, \underline{\rho}_{s'}^{(j)}, \V{o}_{n}, \underline{a}_{ss',n}^{(j)};\V{z}_{n}^{(j)})  
		\mathrm{d} \V{x}_{n} \mathrm{d} \underline{\V{p}}_{s,\mathrm{\sfv}}^{(j)} \mathrm{d} \underline{\rho}_{s}^{(j)} \mathrm{d} \underline{\V{p}}_{s',\mathrm{\sfv}}^{(j)} \mathrm{d} \underline{\rho}_{s'}^{(j)} } 
	\nn \\ &
	+ \underbracea{ 1(\underline{a}_{ss',n}^{(j)}) \Big( \alpha_{s,n}^{(j)} \int \alpha(\underline{\V{p}}_{s',\mathrm{\sfv}}^{(j)}, \underline{r}_{s',n}^{(j)} \rrmv=\rrmv 1, \underline{\rho}_{s'}^{(j)} ) \mathrm{d} \underline{\V{p}}_{s',\mathrm{\sfv}}^{(j)} \mathrm{d} \underline{\rho}_{s'}^{(j)} }_{\hspace*{20mm} \overline{R}_{ss',n}^{(j)}}  
	\nn \\ & 
	\underbraced { + \alpha_{s',n}^{(j)} \int \alpha(\underline{\V{p}}_{s,\mathrm{\sfv}}^{(j)}, \underline{r}_{s,n}^{(j)} \rrmv=\rrmv 1, \underline{\rho}_{s}^{(j)} ) \mathrm{d} \underline{\V{p}}_{s,\mathrm{\sfv}}^{(j)} \mathrm{d} \underline{\rho}_{s}^{(j)} + \alpha_{s,n}^{(j)}\alpha_{s',n}^{(j)}\Big) } \, .
	\label{eq:MeaEvaPSFV2}
\end{align}	

\subsubsection{Measurement Update for Grid Cells}
For the update of the grid cell states, only the messages from the legacy \acp{psfv} are used. The messages $ \tilde{\kappa}_{ss'}^{(j)}(o_{i,n}) $, $ \tilde{\tilde{\kappa}}_{ss'}^{(j)}(o_{i,n}) $ and $ \breve{\kappa}_{ss'}^{(j)}(o_{i,n}) $ with $ (s\rmv,\rmv s') \in \tilde{\mathcal{D}}_{n}^{(j)} $ sent from the factor node $\underline{q}_{\mathrm{P}}(\cdots) $, $\underline{q}_{\mathrm{S}}(\cdots) $ or $\underline{q}_{\mathrm{D}}(\cdots) $ to the cell variables $ o_{i,n} $ are given as follows. If this cell is traversed by the ray, i.e., $ i \in \tilde{\mathcal{Q}}_{ss',n}^{(j)} $,
\begin{align}
	\tilde{\kappa}_{ss'}^{(j)}(o_{i,n}=0) & = \rrmv \rrmv \sum_{\underline{a}_{ss',n}^{(j)} \in \mathcal{M}_{0,n}^{(j)} }  \rrmv \eta(\underline{a}_{ss',n}^{(j)}) \big( R_{ss',n}^{(j)}/\alpha(o_{i,n} = 0) \big) \nn \\ &  
	+ \eta(\underline{a}_{ss',n}^{(j)}=0) \overline{R}_{ss',n}^{(j)} 
	\label{eq:beliefCell1} 
\end{align}
where $\mathcal{M}_{0,n}^{(j)} \triangleq \{0\} \cup \mathcal{M}_{n}^{(j)}$, and $ \tilde{\kappa}_{ss'}^{(j)}(o_{i,n} = 1) = \eta(\underline{a}_{ss',n}^{(j)}=0) \overline{R}_{ss',n}^{(j)} $. If the cell belongs to the hit set, i.e., $ i \in \tilde{\tilde{\mathcal{Q}}}_{ss',n}^{(j)}  $,
\begin{align}
	\tilde{\tilde{\kappa}}_{ss'}^{(j)}(o_{i,n} = 1) & = \rrmv \rrmv \sum_{\underline{a}_{ss',n}^{(j)} \in \mathcal{M}_{0,n}^{(j)} }  \rrmv  \rrmv \eta(\underline{a}_{ss',n}^{(j)}) \big( R_{ss',n}^{(j)}/\alpha(o_{i,n} = 1) \big) \nn \\ &  
	+ \eta(\underline{a}_{ss',n}^{(j)}=0) \overline{R}_{ss',n}^{(j)} \, ,
	\label{eq:beliefCell2} 
\end{align}
and $ \tilde{\tilde{\kappa}}_{ss'}^{(j)}(o_{i,n} = 0) = \eta(\underline{a}_{ss',n}^{(j)}=0) \overline{R}_{ss',n}^{(j)} $. If the cell does not interact with this ray $ i \in \breve{\mathcal{Q}}_{ss',n}^{(j)} $, we still formulate the update to maintain complete statistics, yielding
\begin{align}
	\breve{\kappa}_{ss'}^{(j)}(o_{i,n}) = \rmv\rrmv \hspace{-1mm}  \sum_{\underline{a}_{ss',n}^{(j)} \in \mathcal{M}_{0,n}^{(j)} }  \rmv\rrmv \hspace{-1mm}  \eta(\underline{a}_{ss',n}^{(j)})  R_{ss',n}^{(j)} + \eta(\underline{a}_{ss',n}^{(j)}=0) \overline{R}_{ss',n}^{(j)} \, .
	\label{eq:beliefCell3} 
\end{align}
\subsubsection{Belief Calculation}
With above messages from all \acp{pa}, the beliefs that approximate $ f(o_{i,n}\mid \V{z}_{1:n}) $ are given by
\vspace{-1mm}
\begin{align}
	\tilde{f}\left(o_{i,n}\right) = C_{\mathrm{o}} \alpha(o_{i,n}) \prod_{j=1}^{J} \hspace*{-0.5mm} \prod_{ \substack{ (s\rmv,\rmv s')  \in \tilde{\mathcal{D}}_{n}^{(j)} } } \hspace*{-2.3mm}  \kappa_{ss'}^{(j)}(o_{i,n}) 
	\label{eq:beliefCell} \\[-8mm]\nn
\end{align}
where $\kappa_{ss'}^{(j)}(o_{i,n})$ is the message contributed by path $(s,s')$ to cell $i$. Specifically, $\kappa_{ss'}^{(j)}$ is chosen as $\tilde{\kappa}_{ss',n}^{(j)}$, $\tilde{\tilde{\kappa}}_{ss',n}^{(j)}$, or $\check{\kappa}_{ss',n}^{(j)}$ in \eqref{eq:beliefCell1}-\eqref{eq:beliefCell3} depending on whether cell $i$ belongs to $\tilde{\mathcal{Q}}_{ss',n}^{(j)}$, $\tilde{\tilde{\mathcal{Q}}}_{ss',n}^{(j)}$, or
$\check{\mathcal{Q}}_{ss',n}^{(j)}$, respectively. The normalization constant $ C_{\mathrm{o}} $ ensures that the belief is a valid probability distribution.

\vspace{-2mm}
\subsection{Implementation Aspects}
In particle-based implementations, the evaluation of particle-grid interactions and the calculation of associated statistical updates are computationally intensive. While particle independence enables efficient parallelization, careful aggregation across particles and propagation paths along with normalization is crucial for maintaining numerical stability and probabilistic consistency. In particular, for each propagation path, the relevant statistics are first integrated over all particles before combining information across paths and \acp{pa} in \eqref{eq:beliefCell}.

%% file: InputFiles/Results.tex
\begin{figure*}[t]
	\hspace*{-1mm}
	\begin{minipage}[t]{.32\textwidth}
		\hspace*{0mm}\subfloat[\label{subfig:synEnv}]
		{\hspace*{-2mm}\includegraphics[scale=1]{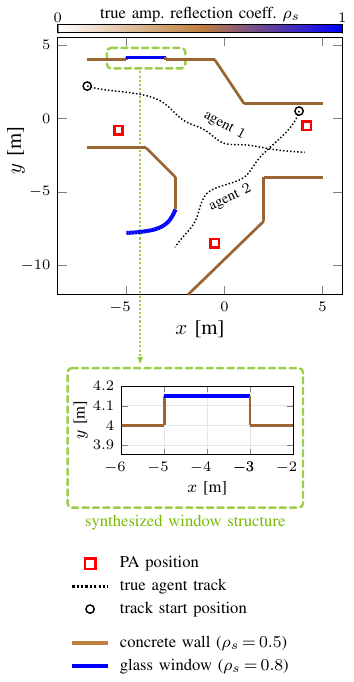}}
	\end{minipage}
	\hspace*{-2.5mm}
	\begin{minipage}[t]{.32\textwidth}
		\vspace*{-118.7mm}
		\hspace*{7mm}\subfloat[\label{subfig:AmpFactor_agent1}]
		{\hspace*{-7mm}\includegraphics[scale=1]{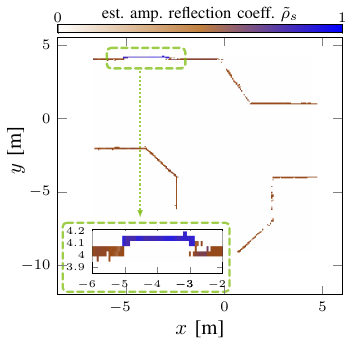}}\vspace*{0mm} \\ [-3mm]
		\hspace*{9mm}\subfloat[\label{subfig:AmpFactor_agent2}]
		{\hspace*{-9mm}\includegraphics[scale=1]{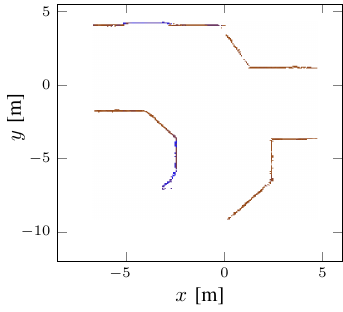}}\vspace*{0mm}
	\end{minipage}
	\hspace*{0mm}
	\begin{minipage}[t]{.32\textwidth}
		\vspace*{-118.7mm}
		\hspace*{7mm}\subfloat[\label{subfig:OccProb_agent1}]
		{\hspace*{-7mm}\includegraphics[scale=1]{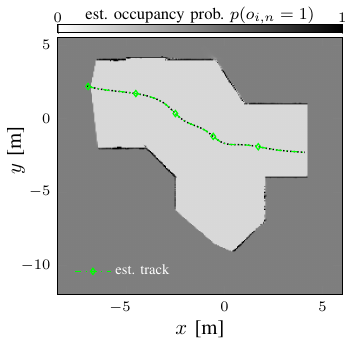}}\vspace*{0mm} \\ [-3mm]
		\hspace*{9mm}\subfloat[\label{subfig:OccProb_agent2}]
		{\hspace*{-9mm}\includegraphics[scale=1]{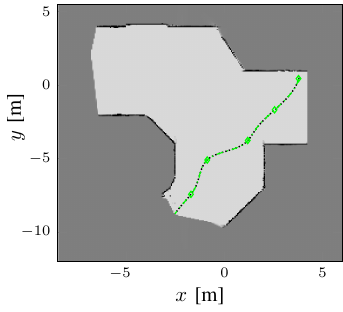}}\vspace*{0mm} 
	\end{minipage}
	\caption{Performance results using synthetic \ac{dmimo} measurements. (a) Geometrical depiction of the simulation environment and setup. (b) and (d) show the localization and mapping results for agent~$ 1 $. (c) and (e) present the results for agent~$ 2 $ using the established occupancy grid map from agent~$ 1 $ in (d) as prior for $ p(\V{o}_{0}) $.}
	\label{fig:results}
	\vspace{1mm}
\end{figure*}

The performance of the proposed algorithm is validated using synthetic radio measurements generated in a challenging environment, as shown in Fig.~\ref{subfig:synEnv}, which could represent a complex indoor environment or an outdoor urban street canyon. This environment consists of three \acp{pa} and features geometrically complex and nonconvex structures with concrete walls and glass windows, characterized by reflection coefficients of $ \rho_{s} = 0.5$ and $ \rho_{s} = 0.8 $, respectively. Agent~$ 1 $ and agent~$ 2 $ explore the \ac{roi} sequentially. 

\vspace*{-1mm}
\subsection{Analysis Setup}
The state-transition \ac{pdf} $f(\V{{\beta}}_{n}|\V{{\beta}}_{n-1}) $ of the agent is defined by a linear near constant-velocity motion model \cite{BarShalom_AlgorithmHandbook} as $ \V{{\beta}}_{n} = \V{F}\V{{\beta}}_{n-1} + \V{\Gamma}\V{\nu}_{n} $, where the transition matrices $\V{F} \in \mathbb{R}^{4\times4} $ and $ \V{\Gamma} \in \mathbb{R}^{4\times2} $ are chosen as in \cite{BarShalom_AlgorithmHandbook}. The driving process $ \V{\nu}_{n} \in \mathbb{R}^{2\times1} $ is independent and identically distributed (iid) over time $ n $, zero-mean and Gaussian with covariance matrix $ \sigma_{\mathrm{\nu}}^2\mathbf{I}_{2} $. The state-transition \ac{pdf} of the agent's orientation is given as $ {\Delta\varphi}_{n} = {\Delta\varphi}_{n-1} + {\epsilon_{\mathrm{\varphi},n}} $ where the noise ${\epsilon_{\mathrm{\varphi},n}}$ is zero-mean and Gaussian with variance ${\sigma^2_{\mathrm{\varphi}}}$. The state-transition \ac{pdf} of the positions of legacy \ac{psfv} $\underline{\V{p}}_{s,\mathrm{\sfv}}$ is chosen to be $ \underline{\V{p}}_{s,\mathrm{\sfv}} = \V{p}_{s,\mathrm{\sfv}} + {\V{\epsilon}}_{s,n} $, where the noise ${\V{\epsilon}}_{s,n}$ is iid, zero-mean Gaussian across $ j $ and $ n $, with variance ${\underline{\sigma}_{\mathrm{p}}}_{s}^2\mathbf{I}_{2}$. The state-transition \ac{pdf} of the surface reflection coefficient $\underline{\rv{{\rho}}}_{s,n}$ is chosen to be $ \underline{\rho}_{s,n} = \rho_{s} + {\epsilon_{\mathrm{{\rho}}}}_{s} $, where the noise ${\epsilon_{\mathrm{{\rho}}}}_{s}$ is iid across $ s $, zero-mean, and Gaussian with variance $\sigma_{\mathrm{{\rho}}}^2$. The state-transition model of the grid cell states is given as $ {o}_{i,n} = {o}_{i,n-1}$. We performed the simulation with the following parameters: the detection threshold for normalized amplitude measurements $u_{\mathrm{de}} =6$\,dB, the acceleration noise \ac{std} $ \sigma_{\mathrm{\nu}} = 9\cdot10^{-4}\,$$ \mathrm{m}/\mathrm{s}^2 $, the orientation noise \ac{std} ${\sigma_{\mathrm{\varphi}}} = 2^{\circ}$, the regularization noise \ac{std} for \ac{psfv} positions ${\underline{\sigma}_{\mathrm{p}}}_{s} = 0.01\,$m, the \ac{std} for the reflection coefficient $ \sigma_{\mathrm{{\rho}}} = 0.01$.

The samples for the initial agent state are drawn from a $5$-D uniform distribution centered at $ [\V{p}_{0}^{\mathrm{T}}\iist 0 \iist 0 \iist {\Delta\varphi}_{0}]^{\mathrm{T}} $ where $ \V{p}_{0} $ and ${\Delta\varphi}_{0}$ are the true agent start position and orientation, and the support of position, velocity and orientation components are $ [-0.5,0.5]\,$m, $ [-0.01,0.01]\,$m/s and $ [-5^{\circ},5^{\circ}]$, respectively. The samples representing the initial state of the new \acp{psfv} $\V{p}_{\mathrm{\sfv}}^{\mathrm{ini}}$ are drawn from the \ac{pdf} $f(\V{p}_{\mathrm{\sfv}}^{\mathrm{ini}}) $ which is uniform in \ac{roi}, i.e., $[-15\mathrm{m},15\mathrm{m}] \times[-15\mathrm{m},15\mathrm{m}] $ centered at the coordinate center. For any new \acp{psfv}, it is assumed that the initial state of the reflection coefficient $\rho_{m}^{\mathrm{ini}}$ is uniform on $ [0.1, 0.9] $. The mean number of new \acp{psfv} is $\mu_{\mathrm{n}} = 0.05$, the survival probability is $p_{\mathrm{s}} = 0.99$, the detection and pruning thresholds are $ p_\mathrm{de} =0.5 $ and $p_\mathrm{pr} = 0.1$, and the \acp{pdf} of the states of the random variables are represented by $ 13000 $ particles each. 

\vspace*{-1mm}
\subsection{Synthetic Measurements}
We synthetically generate the measurements $ \V{z}_{m,n}^{(j)} $ according to the scenario shown in Fig.~\ref{subfig:synEnv}, where a \ac{dmimo} system operating at $f_{\mathrm{c}} = 6$\,GHz with a bandwidth of $1$\,GHz is used. For the three \acp{pa} and two mobile agents, the same $ 5 \times 5 $ uniform rectangular array with an inter-element spacing of $\lambda/4$ is used. Over time steps, time-varying distances, \acp{aoa}, \acp{aod}, and normalized amplitudes were synthesized for propagation paths experiencing up to double surface reflections. The amplitude of each path is assumed to follow free-space pathloss and is attenuated by the reflection coefficient $ {\rho}_{s} $ after each surface reflection. Noisy measurements that account for non-ideal channel estimation and detection are obtained by adding noise to the true path parameters and stacked in the vector $\V{z}_{n}^{(j)}$\!, where the noise statistics are determined via Fisher information matrix \cite{Thomas_Asilomar2018, LeitingerICC2019, XuhongTWC2022} with $\mathrm{SNR}_{\mathrm{1m}}=30$\,dB, the output \ac{snr} at $1$\,m from each \ac{pa}. In addition, the \ac{fa} measurements are further added to $\V{z}_{n}^{(j)}$\! with mean number $ \mu_{\mathrm{fa}}=1 $. Agent~$ 1 $ follows a trajectory of $ 250 $ time steps and agent~$ 2 $ of $ 230 $ time steps, with a step size of $ 5 $\,cm. For modeling the occupancy probability of the environment, the \ac{roi} that spans $ 14.5$\,m$\times17.5 $\,m is divided into square cells with side lengths of $ 0.06 $\,m, resulting in $ Q = 70664 $ grid cells. All cells are initialized with a uniform prior with $ p(o_{i,0}=1) = 0.5 \ \forall i \in \mathcal{Q} $ representing an unknown state that has not been explored, and a lower threshold of $ 0.15 $ is applied to facilitate rapid detection of objects in the estimated free area. Based on the \ac{psfv} and agent states, the path states are obtained and represented by $ 13000 $ particles (i.e., individual rays). Parallel ray casting is performed to determine ray--cell interactions, implemented in Taichi (Python) \cite{taichi_repo}.

\vspace{-2mm}
\subsection{Results}
\vspace{-0.5mm}
Using synthetic measurements, we evaluate (i) localization performance, (ii) mapping performance assessed via reconstructed geometry, estimated material properties (captured by the reflection coefficients), and the ability to incrementally expand and update an existing map, and (iii) ``soft'' ray tracing, i.e., tracking specular propagation paths along with their geometric parameters, amplitudes, and associated variances. Starting from a uniform prior (i.e., without environmental knowledge), agent~$ 1 $ first explores the \ac{roi} to construct an initial map. Agent~$ 2 $ then uses this established map to continue exploration and refine reconstruction. The estimated reflection coefficients $ \tilde{\rho}_{s} $ are shown in Figs.~\ref{subfig:AmpFactor_agent1} and \ref{subfig:AmpFactor_agent2} for agents~$ 1 $ and $ 2 $, respectively, where $ \tilde{\rho}_{s} $ is projected onto grid cells with occupancy probabilities exceeding the initial value of $ 0.5 $. The resulting \ac{rf}-related material maps, together with the occupancy reconstructions in Figs.~\ref{subfig:OccProb_agent1} and \ref{subfig:OccProb_agent2}, accurately recover the geometry of the environment, including fine structures such as the window and the curved glass wall, and their \ac{rf}-related properties. Moreover, Figs.~\ref{subfig:OccProb_agent1} and \ref{subfig:OccProb_agent2} demonstrate the accurate localization performance of the proposed algorithm, achieving average position and orientation errors of $ 0.6 $\,cm and $ 0.27^{\circ} $ for agent~1 and $ 0.7 $\,cm and $ 0.33^{\circ} $ for agent~2.  

\begin{figure}[!t]
	\captionsetup[subfigure]{labelformat=empty, labelsep=none} 
	\centering
	\vspace*{-4mm}
	\hspace*{6mm}\subfloat[]
	{\hspace*{-6mm}\includegraphics[scale=0.98]{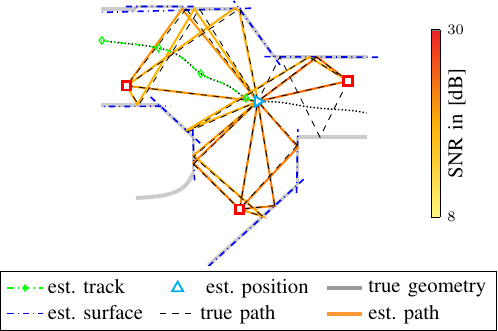}}\\[-1.5mm]
	\caption{Results of a simulation run for agent $ 1 $. The estimated surfaces, propagation paths and agent track are shown for time $ n=173 $. The estimated surfaces are computed using the \ac{mmse} estimates of the detected \acp{sfv}. Estimated propagation paths are obtained by connecting the \ac{mmse} estimates of the agent position, interaction points on the estimated surfaces and \acp{pa}, and compared with the true visible paths. The color of each estimated path represents its \ac{snr} estimates, i.e., the square of normalized-amplitude estimates.}	
	\label{fig:synResultAllPA}
	\vspace{-1mm}
\end{figure}

The ``soft'' ray tracing results in Fig.~\ref{fig:synResultAllPA} illustrate the path-level output of the proposed algorithm using measurements for agent~$1$. The estimated surfaces, obtained from the \ac{mmse} estimates of the \acp{psfv}, closely match the ground-truth geometry, and the specular propagation paths are accurately detected and estimated, including their amplitudes. Unlike previous work \cite{MVASLAM_TWC2025Arxiv} that modeled and inferred per-path existence and amplitude states, the proposed approach infers path geometry and amplitudes directly from the \ac{psfv} estimates and provides Fisher information-based variances. Such path-level information is useful beyond sensing, e.g., to support communications, facilitate channel prediction, beamforming and beam alignment.

%% file: InputFiles/Conclusion.tex
This paper presented a probabilistic occupancy grid framework for radio-based \ac{slam}, enabling the reconstruction of complex environmental geometries together with their radio-related properties. The proposed approach jointly estimates the occupancy states of the grid cells, the reflection coefficients characterizing material properties, and the agent positions. Simulation results demonstrate accurate geometry and material reconstruction, high localization accuracy, and the ability to augment an existing map through subsequent exploration. Possible future research directions include extending the proposed framework to more realistic signal models and diverse feature models beyond planar surfaces.